\documentclass[]{aa}  

\usepackage{graphicx}
\usepackage{txfonts}
\usepackage{xcolor}
\usepackage{soul}

\begin{document}

   \title{X-ray photodesorption of complex organic molecules in protoplanetary disks}

   \subtitle{I. Acetonitrile CH$_3$CN}

   \author{R. Basalgète
          \inst{1}
          \and
          D. Torres-Díaz
          \inst{1, 2}
          \and
          A. Lafosse
          \inst{2}
          \and
          L. Amiaud
          \inst{2}
          \and
          G. Féraud
          \inst{1}
          \and
          P. Jeseck
          \inst{1}
          \and
          L. Philippe
          \inst{1}
          \and
          X. Michaut
          \inst{1}
          \and
          J.-H. Fillion
          \inst{1}
          \and
          M. Bertin
          \inst{1}
          }

   \institute{Sorbonne Université, Observatoire de Paris, Université PSL, CNRS, LERMA, F-75005 Paris, France\\
   \email{romain.basalgete@sorbonne-universite.fr}
    \and
    Université Paris-Saclay, CNRS, ISMO, 91405 Orsay, France
    }

   \date{Received xx; accepted xx}

 
  \abstract
   {X-rays emitted from pre-main-sequence stars at the center of protoplanetary disks can induce nonthermal desorption from interstellar ices populating the cold regions of the disk. This process, known as X-ray photodesorption, needs to be quantified for complex organic molecules (COMs), including acetonitrile CH$_3$CN, which has been detected in several disks.}
   {The purpose of this work is to  experimentally estimate the X-ray photodesorption yields of neutral species from pure CH$_3$CN ices and from interstellar ice analogs for which CH$_3$CN is mixed either in a CO-dominated ice or in a H$_2$O-dominated ice.  }
   {The ices, grown in an ultrahigh vacuum chamber, were irradiated at 15 K by soft X-rays from synchrotron light (SOLEIL synchrotron) in the N K edge region (395 - 420 eV) and in the O K edge region (530 - 555 eV). X-ray photodesorption was probed in the gas phase via quadrupole mass spectrometry by monitoring the changes in the mass signals due to the X-ray irradiation of the ices. X-ray photodesorption yields were derived from the mass signals and were extrapolated to higher X-ray energies in order to provide astrophysical yields adapted to astrochemical models. }
   {X-ray photodesorption of the intact CH$_3$CN is detected from pure CH$_3$CN ices and from mixed $^{13}$CO:CH$_3$CN ices, with an experimental yield of about 5 $\times$ 10$^{-4}$ molecules.photon$^{-1}$ at 560 eV. When mixed in H$_2$O-dominated ices, X-ray photodesorption of the intact CH$_3$CN at 560 eV is below its detection limit, which is 10$^{-4}$ molecules.photon$^{-1}$. Yields associated with the desorption of HCN, CH$_4$ , and CH$_3$ are also provided. The derived astrophysical yields significantly depend on the local conditions expected in protoplanetary disks, that is, on the ice composition and on the local X-ray irradiation spectrum. They vary from $\sim$ 10$^{-4}$ to $\sim$ 10$^{-6}$ molecules.photon$^{-1}$ for the X-ray photodesorption of intact CH$_3$CN from CO-dominated ices. Only upper limits varying from $\sim$ 5 $\times$ 10$^{-5}$ to $\sim$ 5 $\times$ 10$^{-7}$ molecules.photon$^{-1}$ could be derived for the X-ray photodesorption of intact CH$_3$CN from H$_2$O-dominated ices.   }
   {X-ray photodesorption of intact CH$_3$CN from interstellar ices might in part explain the abundances of CH$_3$CN observed in protoplanetary disks. The desorption efficiency is expected to vary with the local physical conditions, hence with the disk region considered. }


   \maketitle
%

\section{Introduction}
The detection of complex organic molecules (COMs) in protoplanetary disks at the very early stages of planet formation raises the question of their role in the emergence of life in nascent planets via prebiotic chemistry. Gaseous acetonitrile CH$_3$CN, one COM, is detected in several disks \citep{oberg_comet-like_2015, bergner_survey_2018, loomis_distribution_2018}. Its formation pathways in the interstellar medium (ISM) include both gas-phase reactions and energetic or nonenergetic ice chemistry. However, disk modeling studies that include gas-phase pathways alone fail to reproduce the observed abundances of CH$_3$CN \citep{oberg_comet-like_2015, loomis_distribution_2018}. Instead, the models suggest an ice formation route and a subsequent delivery of CH$_3$CN to the gas phase via nonthermal desorption processes. In particular, it is deduced from the models that gas-phase CH$_3$CN should be dominantly present in the upper layers of the observed disks \citep{oberg_comet-like_2015, loomis_distribution_2018}, where photons emitted from the pre-main-sequence (PMS) star irradiate the ices. It is therefore expected that photon-induced desorption, known as photodesorption, should play an important role in explaining gas-phase CH$_3$CN in disks. As mentioned in \cite{oberg_comet-like_2015} and \cite{loomis_distribution_2018}, these photodesorption processes are poorly constrained experimentally.  
\\\\
Recently, vacuum ultraviolet (VUV) photodesorption in the 7 - 13.6 eV range of intact CH$_3$CN from interstellar ice analogs has been experimentally demonstrated \citep{basalgete_photodesorption_2021}. However, the derived photodesorption yields ($\sim$ 10$^{-5}$ molecules.photon$^{-1}$) are two orders of magnitude lower than the yield that was used to explain the column density of the observed CH$_3$CN by disk modeling \citep{loomis_distribution_2018}. This may indicate that nonthermal desorption processes other than VUV photodesorption could be at play in protoplanetary disks. For instance, PMS stars can be strong X-ray emitters \citep{gudel_x-ray_2009, testa_x-ray_2010, feigelson_x-ray_2010}, and laboratory astrophysics experiments conducted in recent years have shown that X-rays can induce desorption from interstellar ice analogs \citep{dupuy_x-ray_2018, jimenez-escobar_x-ray_2018, ciaravella_x-ray_2020, dupuy_x-ray_2021, basalgete_complex_2021, basalgete_complex_2021-b}. Additionally, in a recent modeling study of \cite{notsu_x-ray-induced_2021}, it has been shown that X-ray photodesorption can have a significant influence on the gas-phase abundances of water outside the water snowlines of disks. This further encourages additional experimental studies of X-ray photodesorption from interstellar ices.
\\\\
In this study, we experimentally quantify X-ray photodesorption of neutral species from CH$_3$CN-containing ices. X-ray photodesorption is studied as a function of the ice composition, first from pure ices of acetonitrile, and then from interstellar ice analogs for which CH$_3$CN is mixed in CO-dominated or H$_2$O-dominated ices. These mixed ices serve as model ices representing different cold regions of protoplanetary disks, namely the regions outside the H$_2$O or the CO snowlines where the surface of the ice is expected to be mainly composed of H$_2$O or CO, respectively, but can also contain small quantities of CH$_3$CN. We restrict the studies to ices irradiated at 15 K for different mixtures in order to understand the effect of the ice composition alone, without the effect of the ice temperature, which varies with the disk region that is considered. The studies are conducted in the soft X-ray range, on the SEXTANTS beam line of the SOLEIL synchrotron facility. Two energy ranges were selected: (1) the 395 - 420 eV range, referred to as the N K edge region, where the photoabsorption is dominated by N-bearing species, that is, CH$_3$CN, and (2) the 525 - 560 eV range, referred to as the O K edge region, where the photoabsorption is dominated by O-bearing species, that is, H$_2$O or CO. Consequently, selective photoexcitation of CH$_3$CN, H$_2$O, or CO enables us to study possible indirect desorption mechanisms that have been highlighted in previous studies \citep{basalgete_2022}. X-ray photodesorption yields extrapolated to the 0.4 - 10 keV range and averaged over different attenuated X-ray emission spectra of PMS stars, referred to as astrophysical yields, were derived in order to facilitate the implementation of X-ray photodesorption in astrochemical models. Section \ref{sec:exper} describes the experimental procedure and the derivation of the yields. In Section \ref{sec:results} we present the results, and their astrophysical implications are discussed in Section \ref{sec:astro}. This is paper I of an experimental work dedicated to the study of the X-ray photodesorption of COMs from interstellar ice analogs. Paper II studies the X-ray photodesorption of formic acid HCOOH.

\section{Experimental procedure}\label{sec:exper}
\subsection{Ice deposition, TEY, and synchrotron beam line}
Experiments were conducted using the surface processes and ices (SPICES) setup. It consists of an ultrahigh vacuum chamber (UHV) with a base pressure of $\sim$ 10$^{-10}$ Torr, equipped with a quadrupole mass spectrometer (QMS). At the center of the chamber, a rotatable copper substrate (polycrystalline oxygen-free high-conductivity copper) is mounted on a sample holder that can be cooled down to 15 K by a closed-cycle helium cryostat. The ices are formed on the substrate by injecting gas-phase molecules in the chamber via a tube that can be positioned a few millimeters in front of the substrate surface. Different injection gas lines enable us to deposit binary mixed ices, with dilution ratios that are controlled by adjusting the partial pressure associated with each species during deposition. Isotopologs are used to facilitate the analysis of the mass spectrometer data. Pure acetonitrile $^{12}$CH$_3^{12}$C$^{14}$N (99\% purity, Sigma-Aldrich) and $^{12}$CH$_3^{13}$C$^{15}$N (99\% isotopic purity, Sigma-Aldrich) ices were deposited and irradiated at 15 K. Mixed $^{13}$CO:CH$_3$CN ices ($^{13}$CO from 99\% $^{13}$C purity Eurisotop) were deposited and irradiated at 15 K. Mixed H$_2$O:CH$_3$CN ices (H$_2$O from liquid chromatography standard Fluka) were deposited at 90 K, cooled down to 15 K and irradiated at 15 K. This ensured that the resulting water ice is in its compact amorphous phase, referred to as compact amorphous solid water (c-ASW). The thickness of the grown ices is expressed in monolayers (ML), equivalent to a surface density of $\sim$ 10$^{15}$ molecules.cm$^{-2}$. Temperature-programmed desorption (TPD) experiments conducted prior to the presented studies enabled us to control the number of ML deposited with a precision of about 10\% (see, e.g., \cite{bertin_nitrile_2017} for TPD of acetonitrile). 
\\\\
The substrate was electrically insulated from the sample holder by a Kapton foil. This enabled the measurement of the drain current generated by the escape of electrons from the ice into the vacuum after X-ray absorption. From this current, we derived the total electron yield (TEY), expressed in electrons per incident photon (e$^-$.photon$^{-1}$ for more simplicity), and measured as a function of the incident photon energy. The TEY is sensitive to the changes in the molecular composition near the ice surface with the ongoing irradiation, that is, with the photon fluence (expressed in photons.cm$^{-2}$), and it can be assimilated to the X-ray absorption spectrum of the studied ices. The ice depth probed by the TEY measurements is estimated to be a few tens of ML based on studies of water ice \citep{TIMNEANU_2004} and of CO/N$_2$ ices \citep{basalgete_2022}. 
\\\\
X-rays from the SEXTANTS beam line of the SOLEIL synchrotron facility at Saint-Aubin, France \citep{Sacchi_2013}, were routed to the UHV chamber to irradiate the grown ices. Photons in the N and O K edge regions (395 - 420 eV and 525 - 560 eV, respectively) were used with different spectral width (namely 1.2 eV or 90 meV) and with a flux varying from 10$^{12}$ to 10$^{13}$ photons.s$^{-1}$, the latter was measured by a calibrated silicon photodiode mounted on the beam line. The beam was sent at a 47$^\circ$ incidence relative to the normal of the substrate surface, and the spot area at the surface was $\sim$ 0.1 cm$^2$. The calibration of the energy scale was performed similarly to what is described in \cite{basalgete_2022}. Namely, in the N K edge region, a TEY was measured on a pure N$_2$ ice at 15 K, and the TEY feature corresponding to the N 1s $\rightarrow \pi^* (v^{\prime} = 0)$ transition of N$_2$ was set to 400.868 eV according to \cite{chen_k_1989}. In the O K edge region, a TEY was measured on a pure CO ice at 15 K, and the TEY feature corresponding to the O 1s $\rightarrow \pi^*$ transition of CO was centered at 534.4 eV according to \cite{jugnet_high-resolution_1984}.

\subsection{Derivation of the X-ray photodesorption yields}

The X-ray photodesorption of neutral species was monitored in the gas phase of the UHV chamber during the X-ray irradiation of the ices and by means of the QMS equipped with an electron-impact (at 70 eV) ionization stage. The desorption intensities $I_X(E)$ associated with a desorbing neutral species $X$ for a photon energy $E$ were derived by following the m/z signals of the QMS during the X-ray irradiation. Irradiation at fixed energy for a few tens of seconds results in a sudden increase and decrease in the mass signals that is associated with X-ray photodesorption. $I_X(E)$ was then computed as the height of the signal increase in that case. Examples of these QMS signals are presented in Appendix \ref{App:app_A} in Figure \ref{fig:fig_raw_data_1} for the mass signals m/z 27 from a pure CH$_3^{12}$C$^{14}$N ice and m/z 41 from a mixed $^{13}$CO:CH$_3$CN (10:1) ice. The QMS signals can also be monitored by continuously scanning the incident photon energy, resulting in signals similar to what is displayed in Figure \ref{fig:fig_raw_data_2} of Appendix \ref{App:app_A}. In this case, the timescale was converted into an energy scale and the background level (mass signal without irradiation) was subtracted to derive $I_X(E)$. After the attribution of the m/z channels to desorbing neutral species (see Section \ref{sec:result_pure} and \ref{sec:result_mixed}), the intensities $I_X(E)$ were corrected for the fragmentation of these attributed species due to their ionization by electron impact. The fragmentation patterns were taken from the NIST database
\citep{NIST_chemistry_webbook}. The resulting intensities were then converted into X-ray photodesorption yields $\Gamma_X(E)$, expressed in molecules desorbed per incident photon (simplified to molecules.photon$^{-1}$ in this study), using equation~\ref{eq:yields},
\begin{equation}
    \Gamma_X(E) = k_X \frac{I_X(E)}{\phi(E)}
    \label{eq:yields}
,\end{equation}
where $\phi(E)$ is the photon flux at $E,$ and $k_X$ is a conversion factor associated with the neutral species $X$. The coefficient $k_X$ was calibrated on N$_2$. k$_{N_2}$ relates the QMS current to a calibrated number of N$_2$ molecules desorbed during TPD experiments (see \cite{basalgete_2022} for more details of the calibration procedure). The factor $k_X$ associated with other neutral species was derived from $k_{N_2}$ by taking into account (i) the relative differences in the electron-impact ionization cross sections between N$_2$ and the species $X$ and (ii) the differences in the QMS apparatus function between m/z(N$_2$) and m/z(X). Electron-impact ionization cross sections were taken from the literature for CH$_3$CN \citep{zhou_total_2019}, HCN \citep{pandya_electron_2012}, CH$_4$ \citep{tian_cross_1998} and CH$_3$ \citep{Tarnovsky_1996}.

\subsection{Extrapolation to higher energies. Astrophysical yields}

X-ray photodesorption yields were derived in the soft X-ray range (< 600 eV), whereas X-rays emitted from PMS stars at the center of protoplanetary disks range from 0.1 to 10 keV. We then derived the X-ray photodesorption yields $\Gamma_{astro}$ for mixed ices averaged in the 0.4 - 10 keV range by (i) extrapolating the experimental yields $\Gamma_X$ up to 10 keV and (ii) considering the X-ray emission spectrum $\phi_{local}$ of a typical T-Tauri star (from \cite{nomura_molecular_2007}), which we attenuated by using the photoelectric cross section of gas and dust in a typical T-Tauri protoplanetary disk (from \cite{bethell_photoelectric_2011}). This resulted in the following formula:
\begin{equation}
\label{eq:yield_astro}
    \Gamma_{astro} = \frac{\int \Gamma_X(E) \ \phi_{local}(E) \ dE}{\int \phi_{local}(E) \ dE}
.\end{equation}
The attenuated X-ray emission spectra depend on the column density of gas and dust traversed by X-rays and are displayed in Figure \ref{fig:fig_local_phi} of Appendix \ref{App:astro}. The experimental yields $\Gamma_X$ were extrapolated up to 10 keV by assuming that (i) the X-ray photodesorption yields follow the X-ray absorption profile of the ices, as shown in Section \ref{sec:results} for the N and O K edge regions, and (ii) the X-ray absorption of the ices above 560 eV follows the gas-phase core O 1s ionization cross section, which is similar for H$_2$O and CO and was taken from \cite{Berkowitz:1087021}. Examples of extrapolated yields are given in Figure \ref{fig:fig_extrapol} of Appendix \ref{App:astro}. We assumed that the X-ray photodesorption yields per absorbed photon do not depend on the photon energy, as suggested in \cite{dupuy_x-ray_2018} and \cite{jimenez-escobar_x-ray_2018}. The values of the yields in units of absorbed photons are also provided in Section \ref{sec:astro} based on a similar method as in \cite{basalgete_2022}, assuming that up to 30 ML of the ice contribute to the desorption, and without taking the dilution of CH$_3$CN into account. These yields can easily be extrapolated to environments other than protoplanetary disks.

\section{Results}\label{sec:results}
\subsection{TEYs with photon fluence}\label{sec:TEY}

\begin{figure*}
   \centering
   \includegraphics[scale=0.6]{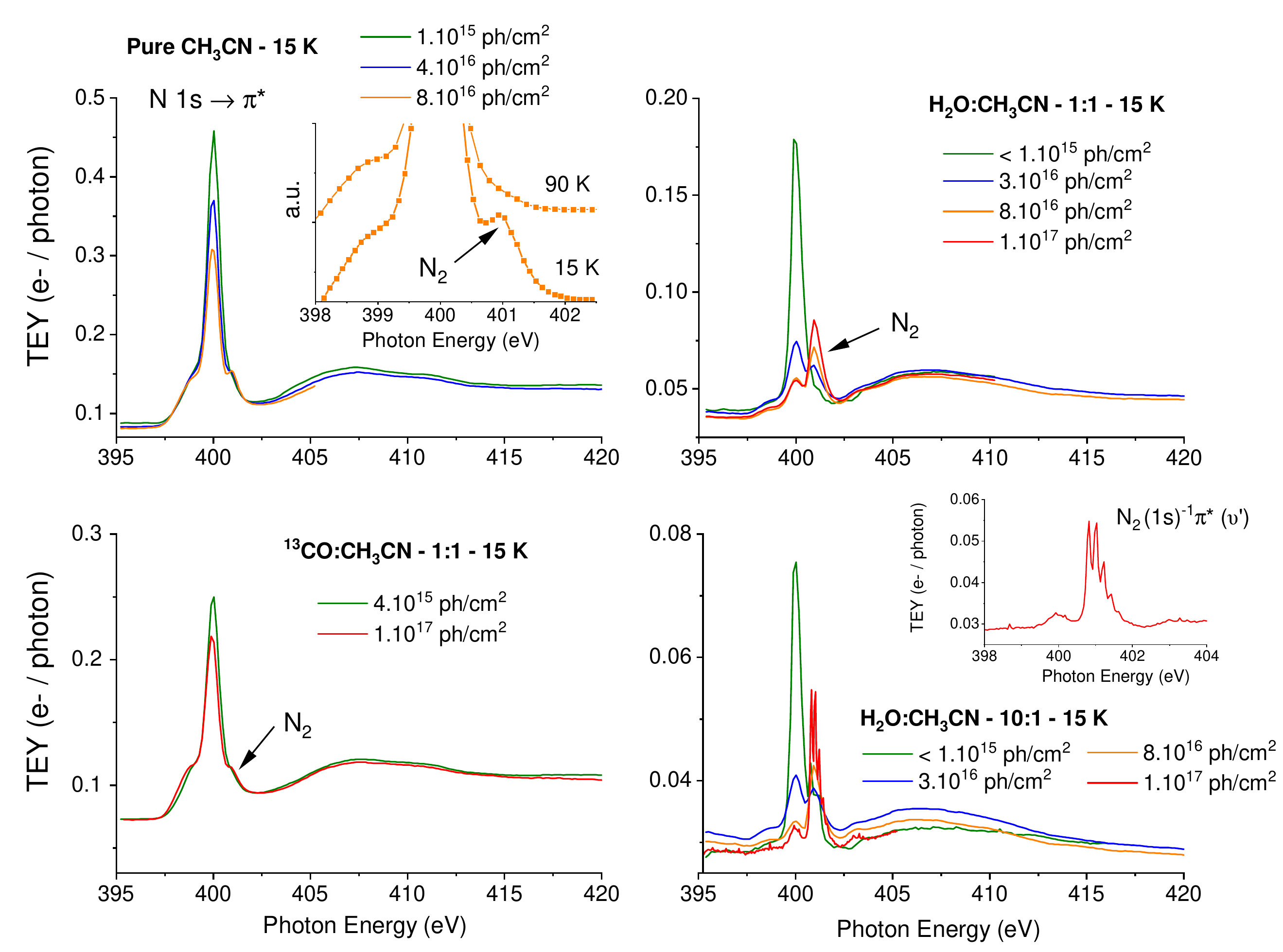}
   \caption{TEYs in the N K edge region of a pure CH$_3$CN ice at 15 K (top left panel; the inset shows the region near the N 1s $\rightarrow \pi^*$ resonance for an ice irradiated at 15 K and 90 K for the lower and upper curve, respectively; these curves are shifted vertically for more clarity), of a mixed H$_2$O:CH$_3$CN ice irradiated at 15 K with a dilution ratio of 1:1 and 10:1 (top and bottom right panel, respectively), and of a mixed $^{13}$CO:CH$_3$CN ice irradiated at 15 K with a dilution ratio of 1:1 (bottom left panel). The photon fluence received by the ice before each TEY measurement is also displayed. The spectral width of the beam was set to 1.2 eV for all the TEY measurements, except for the one corresponding to the red curve in the bottom right panel, for a H$_2$O:CH$_3$CN ice having received a photon fluence of 1$\times$10$^{17}$ photons.cm$^{-2}$ and for which the spectral width was 90 meV. The inset in the bottom right panel zooms into the TEY measured on the H$_2$O:CH$_3$CN (10:1) ice for a photon fluence of 10$^{17}$ photons.cm$^{-2}$ , where the vibrational structure of the core hole state of N$_2$ formed near the ice surface can be seen near 401 eV. The ices have a total thickness of $\sim$ 100 ML.}
              \label{fig:tey_ch3cn_N}%
    \end{figure*}

The TEYs measured on the studied ices are displayed in Figure \ref{fig:tey_ch3cn_N} for the 395 - 420 eV range. Their evolution with the photon fluence, expressed in photons.cm$^{-2}$, is also shown. Our TEYs of the pure CH$_3$CN ice compare well with that of \cite{parent_core-induced_2000}. In this energy range, the X-ray photoabsorption is dominated by N-bearing species. As discussed in the experimental section, the ice depth probed by a TEY measurement is estimated to be a few tens of ML. The evolution of the TEY features with the photon fluence therefore provides information on the changes in the molecular composition near the ice surface, where "near the ice surface" refers to the first tens of ML of the ice. 
\\\\
We first focus on the TEY feature near 400 eV, which dominates the TEYs for low photon fluences and for each studied ice. It is associated with the N 1s $\rightarrow \pi^*$ core transition of CH$_3$CN in the solid phase. Its intensity decreases with the photon fluence due to the photodissociation of CH$_3$CN. In the case of the pure CH$_3$CN ice and the mixed $^{13}$CO:CH$_3$CN (1 :1) ice (see the left panels of Figure \ref{fig:tey_ch3cn_N}), this feature still dominates the TEY for a high photon fluence ($\sim$ 8 - 10 $\times$ 10$^{16}$ photons.cm$^{-2}$), meaning that CH$_3$CN is still present near the ice surface in a significant amount for such fluences. In the case of the mixed H$_2$O:CH$_3$CN ices (1:1 and 10:1), the behavior of the CH$_3$CN feature is very different, as shown in the right panels of Figure \ref{fig:tey_ch3cn_N}. Its decrease with the photon fluence is much faster than for the other studied ices, and it almost totally disappears for a photon fluence $\sim$ 8 $\times$ 10$^{16}$ photons.cm$^{-2}$. This indicates that the water ice provides reactive species, for instance, the OH radical, that increases the destruction kinetics of CH$_3$CN in the case of the H$_2$O:CH$_3$CN ices compared to the case of the pure CH$_3$CN and the $^{13}$CO:CH$_3$CN ices. This kinetic difference in the consumption of CH$_3$CN has also been observed when irradiating pure CH$_3$CN and mixed H$_2$O:CH$_3$CN ices with UV photons (with a broadband 7 - 10.2 eV hydrogen lamp) at 20 K in the study of \cite{bulak_photolysis_2021}. This behavior does not depend on the photon energy in our experiments because the TEYs displayed in Figure \ref{fig:tey_ch3cn_N} were measured for ices that were irradiated both near the N and the O K edges. This is consistent with the fact that the chemistry is dominated by the secondary low-energy electrons created after X-ray absorption and does not depend on the primary photoexcitation or ionization.

\ \\
Among the possible species that formed during the X-ray irradiation, a new feature that appeared with the photon fluence near 401 eV suggests the accumulation of N$_2$ near the ice surface for each studied ice (at 15 K). This feature can be associated with the N 1s $\rightarrow \pi^*$ core transition of N$_2$, as seen from TEY measurements of pure N$_2$ ices \citep{basalgete_2022} and similar to K-shell photoabsorption studies of gas-phase N$_2$ \citep{chen_k_1989}. The absence of this feature in the TEY of the pure CH$_3$CN ice irradiated at 90 K supports its attribution to N$_2$ formation, as N$_2$ would thermally desorbs at this temperature. In the inset of the bottom right panel of Figure \ref{fig:tey_ch3cn_N}, the red curve clearly and definitively confirms this attribution. When the spectral resolution is high enough (in this case, 90 meV), the vibrational structure of the N (1s)$^{-1} \pi^*$ state of N$_2$ is resolved in the TEY, similarly to what has been observed for pure N$_2$ ice in \cite{basalgete_2022}. Possible N$_2$ contamination from the UHV chamber or from the X-ray beam line that would deposit at the ice surface and significantly contribute to the measured TEY seems unlikely because the base pressure was kept at 10$^{-10}$ Torr during the experiments, excluding significant deposition of contaminants on the ice surface on the experimental timescale. Any possible N$_2$ contamination of the sample was searched for and ruled out, for example, via performing TEY measurements on fresh ices at the N-K edge of N$_2$, and on the bare copper substrate at low temperature. The solid N$_2$ TEY signal was found to be directly correlated to the photon irradiation (it depends on the fluence and photon energy) and also to the amount of condensed acetonitrile deposited for a given fluence condition. This gives us confidence that N$_2$ is indeed photoproduced from the solid CH$_3$CN during irradiation. Surprisingly, N$_2$ is formed upon X-ray irradiation regardless of the ice composition. However, further investigations are needed to assess how its formation pathways and its formation kinetics depend on the ice composition. For the mixed ices, where CH$_3$CN molecules are less likely to be spatially close to each other in the ice, the diffusion of N-bearing radicals and/or the formation of CH$_3$CN clusters or islands during the ice deposition might partly explain the formation of N$_2$. For instance, \cite{Jiménez-Escobar_2022} highlighted that the irradiation of interstellar ices by X-rays can induce the diffusion of species through hundreds of ML.
\\\\
Any photoproducts other than N$_2$ that are formed during the X-ray irradiation of each studied ice do not participate significantly in the photoabsorption in the 395 - 420 eV range because no significant new features other than that of N$_2$ appear in the TEY with the photon fluence. It is clear, however, that we do not have the full picture of the X-ray induced chemistry in the TEY measurements. In the literature, low-energy electron irradiation of pure acetonitrile ices \citep{ipolyi_2007, bass_reactions_2012} suggests the formation of HCN and C$_2$H$_6$. \cite{abdoul_2022} suggested the formation of CH$_3$OH (detected by TPD) in mixed H$_2$O:CH$_3$CN ice irradiated by low-energy electrons. As the CH$_3$OH absorption features overlap with that of H$_2$O in the TEYs near the O K edge, we cannot discuss its formation with our data set. In VUV irradiation experiments of H$_2$O:CH$_3$CN ices, the formation of larger COMs was reported \citep{bulak_photolysis_2021}  even though VUV and
X-ray photochemistry are not necessarily comparable.
\\\\
The TEYs measured near the O K edge for mixed $^{13}$CO${:}$CH$_3$CN and H$_2$O${:}$CH$_3$CN ices are displayed in Appendix \ref{App_B} (Figure \ref{fig:tey_ch3cn_O}). The observed features are similar to the feature corresponding to pure H$_2$O and pure CO ice that was studied in \cite{dupuy_desorption_2020} and \cite{dupuy_x-ray_2021}. The main feature for the $^{13}$CO${:}$CH$_3$CN ice is associated with the O 1s $\rightarrow \pi^*$ transition of $^{13}$CO near 534.4 eV. The features associated with H$_2$O are discussed in more detail in \cite{dupuy_desorption_2020}. Significant modifications of these TEYs with the photon fluence are not observed, meaning that potential photoproducts that formed during the X-ray irradiation of the mixed ices do not significantly participate in the photoabsorption of the ices in the 530 - 555 eV range. 

\subsection{X-ray photodesorption from pure CH$_3$CN ice}\label{sec:result_pure}

\begin{figure}[b!]
\centering
\includegraphics[scale=0.48]{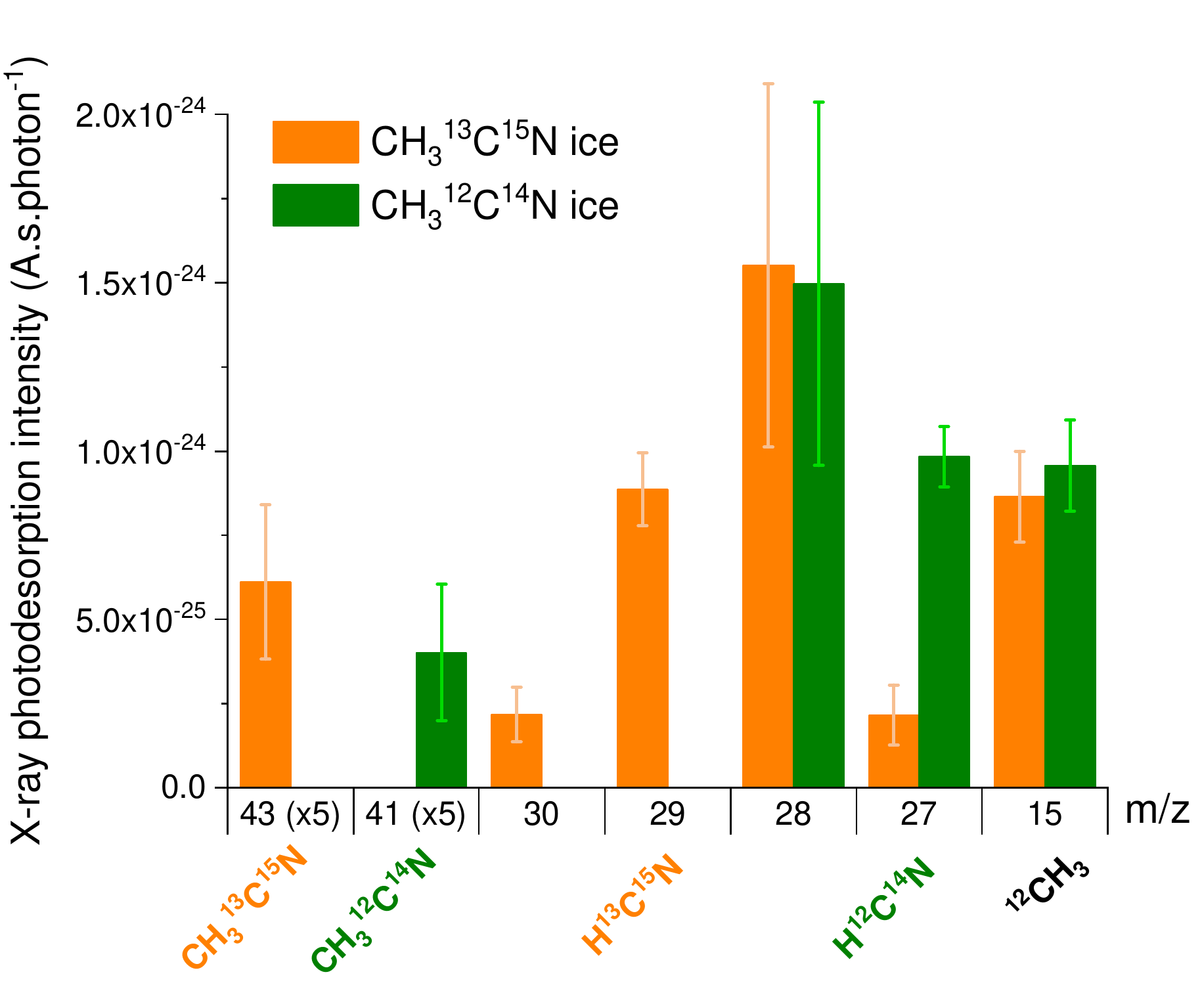} 
\caption{X-ray photodesorption intensities divided by the photon flux (in A.s.photon$^{-1}$) at 420 eV of desorbing species from pure acetonitrile ices. The mass channels monitored during the experiments are indicated on the X-axis. The attribution of these mass channels to desorbing neutral species is discussed in the text. The different colors are associated with a natural CH$_3^{12}$C$^{14}$N (green) or an isotopic CH$_3^{13}$C$^{15}$N (orange) ice, irradiated at 15 K. The signals were obtained for a fluence $<$ 2 $\times$ 10$^{16}$ photons.cm$^{-2}$ , and they are not corrected for any possible fragmentation of desorbing species in the ionization stage of our QMS.}
\label{fig:attrib_mass_ch3cn}
\end{figure}
Pure acetonitrile ices were irradiated at 15 K in the N K edge region (395 - 420 eV). X-ray photodesorption was detected in several mass channels of the QMS. Isotopologs CH$_3^{12}$C$^{14}$N and CH$_3^{13}$C$^{15}$N were used to attribute desorbing neutral species to the mass channels. In Figure \ref{fig:attrib_mass_ch3cn} we display the variations in the desorption intensities (divided by the photon flux) with the isotopolog for pure acetonitrile ices irradiated at 15 K and at 420 eV in the ionization region of the N 1s electron. At this point, the displayed intensities are not corrected for any possible fragmentation pattern of the desorbing species. The desorption intensities of the m/z 41 and 43 from pure CH$_3^{12}$C$^{14}$N and CH$_3^{13}$C$^{15}$N ices, respectively, are similar. This confirms the X-ray photodesorption of the intact acetonitrile molecule from the studied ices. The desorption intensity of the m/z 15 does not change significantly from one isotopolog to the next, indicating the X-ray photodesorption of the methyl group CH$_3$. The possible desorption of CH$_4$ that would contribute to the m/z 15 signal due to its fragmentation at the QMS entrance can be excluded because no desorption signal on the m/z 16 was detected at 420 eV for either isotopolog. Irradiation of pure acetonitrile ices at 12 K by UV photons and 0.8 MeV protons has revealed the formation of CH$_4$ in \cite{HUDSON2004466}, however.
\\\\
For the CH$_3^{13}$C$^{15}$N ice, the desorption intensity of the m/z 29 is $\sim$ 9.0 $\times$ 10$^{-25}$ As.photon$^{-1}$. A similar level of signal is observed on the m/z 27 for the CH$_3^{12}$C$^{14}$N ice. This indicates the X-ray photodesorption of HCN from the pure acetonitrile ices, explaining the signals on the m/z 29 (H$^{13}$C$^{15}$N) for the CH$_3^{13}$C$^{15}$N ice and on the m/z 27 (H$^{12}$C$^{14}$N) for the CH$_3^{12}$C$^{14}$N ice. HCN formation has previously been suggested (by post-irradiation TPD experiments) in low-energy electron irradiation experiments of pure CH$_3$CN ice at 35 K \citep{ipolyi_2007}. Additionally, low-energy electron-stimulated desorption of CH$_2^-$ from pure CH$_3$CN ice at 30 K observed in the study of \cite{bass_reactions_2012} led the authors to suggest the formation of HCN after the dissociative electron attachment (DEA) of CH$_3$CN into CH$_3^-$ and CN followed by H migration to CN. In our experiments, the X-ray induced chemistry is dominated by the cascade of low-energy secondary electrons after X-ray absorption. It is therefore expected that HCN formation as observed by \cite{ipolyi_2007} and \cite{bass_reactions_2012} and subsequent photodesorption can occur. 
{\renewcommand{\arraystretch}{1.5}
\begin{center}
\begin{table}[b!]
\begin{center}
\caption{\label{tab:yield_ch3cn_x_pure}X-ray photodesorption yields in molecules desorbed per incident photon (molecules.photon$^{-1}$) of CH$_3$CN, HCN, and CH$_3$ from pure CH$_3$CN ices irradiated at 15 K and at a photon energy of 420 eV.}
\begin{tabular}{p{2cm}p{4cm}}
 \hline \hline
 Species & Yield  \\\hline
 CH$_3$CN & 5.2 $\pm$ 1.5 $\times$ 10$^{-4}$  \\
 HCN & 2.5 $\pm$ 0.3 $\times$ 10$^{-3}$ \\
 CH$_3$ & 1.3 $\pm$ 0.7 $\times$ 10$^{-3}$\\
 \hline \hline
 \end{tabular}
\end{center}
\tablefoot{The fluence received by the ice before the measurements is $<$ 2 $\times$ 10$^{16}$ photons.cm$^{-2}$}
\end{table}
\end{center}
}
The attribution of the m/z 30 and 28 is not clear due to the large error bars on the m/z 28 and the fact that the m/z 30 was not recorded for the CH$_3^{12}$C$^{14}$N ice. This complicates the interpretation of these signals. Additionally,  several molecules can contribute to these two mass channels. The m/z 30 observed from the CH$_3^{13}$C$^{15}$N ice could correspond to C$_2$H$_6$ and/or $^{15}$N$_2$ desorption. The desorption of C$_2$H$_6$ is supported by the studies of \cite{ipolyi_2007} and \cite{bass_reactions_2012}, where its formation was suggested to occur via reaction between CH$_3$ radicals after DEA of CH$_3$CN into CN$^-$ and CH$_3$. The desorption of N$_2$, which would contribute to the m/z 30 for the CH$_3^{13}$C$^{15}$N ice and to the m/z 28 for the CH$_3^{12}$C$^{14}$N ice, is supported by its formation near the ice surface, as seen in our TEY data (see Figure \ref{fig:tey_ch3cn_N}). As stated in Section \ref{sec:TEY}, blank experiments on fresh ices or on the bare copper substrate allowed us to rule out any possible N$_2$ contamination from the experimental setup or the beam line. We therefore associate any N$_2$ detection with its X-ray induced formation and subsequent desorption from the ice. The fragmentation of desorbing C$_2$H$_6$ at the QMS entrance could also contribute to the m/z 28 observed for both isotopologs. The m/z 28 observed from the CH$_3^{13}$C$^{15}$N ice could also have a contribution from desorbing $^{13}$C$^{15}$N after DEA of CH$_3^{13}$C$^{15}$N, supported by the anion desorption of CH$_3^-$ observed by \cite{bass_reactions_2012}.  Finally, the entanglement is such that we cannot conclude about the attribution of the m/z 28 and 30 from the pure ices. 

\begin{figure}[t!]
\centering
\includegraphics[scale=0.65]{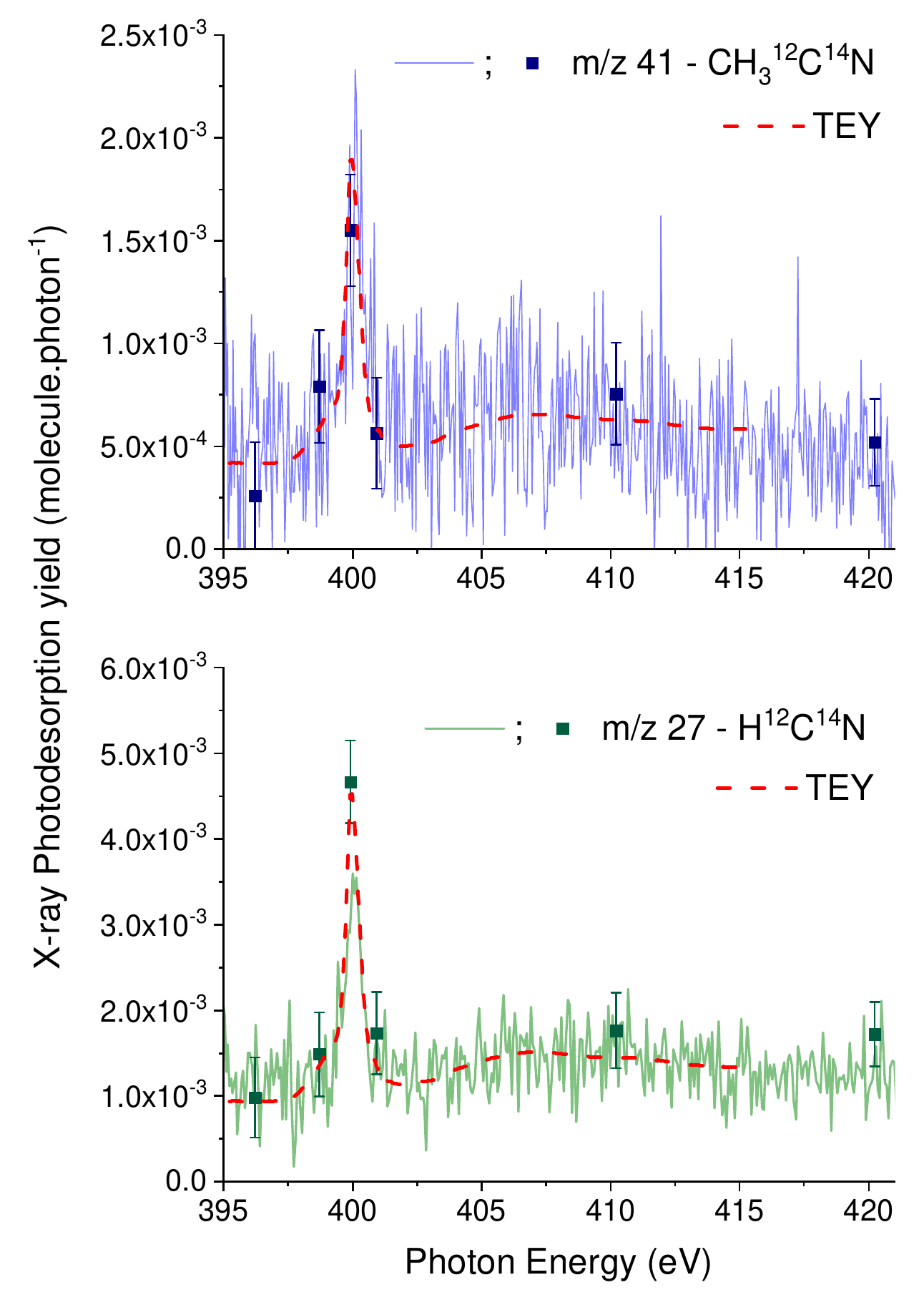} 
\caption{X-ray photodesorption yields of CH$_3^{12}$C$^{14}$N and H$^{12}$C$^{14}$N from a pure CH$_3^{12}$C$^{14}$N ice irradiated at 15 K as a function of the incident photon energy. The solid noisy lines are associated with the desorption yield derived from a continuous irradiation from 395 to 420 eV, whereas the squares with error bars result from the desorption measurements corresponding to irradiation at fixed energies for a few tens of seconds. The TEY measured simultaneously during the continuous irradiation are shown as dashed red lines in arbitrary units.}
\label{fig:fig_psd_pure_ch3cn}
\end{figure}
\ \\
After conversion of the desorption intensities to desorption yields, we display in Table \ref{tab:yield_ch3cn_x_pure} the X-ray photodesorption yields at 420 eV of the identified species (CH$_3$CN, HCN, and CH$_3$) from our experiments on pure acetonitrile ices. The displayed yields are taken as the average on the two isotopologs. We did not correct the HCN yield for the fragmentation of potentially desorbing C$_2$H$_6$, which means that this yield might be overestimated (by $\sim$30\% when we consider that the full intensity measured on the m/z 30 corresponds to C$_2$H$_6$ desorption). The pure acetonitrile ices were also irradiated by varying the photon energy from 395 eV to 420 eV either at fixed energies or by continuously scanning the photon energy. X-ray photodesorption yields were then derived as a function of the photon energy. The resulting photodesorption spectra are shown in Figure \ref{fig:fig_psd_pure_ch3cn} for the desorption of CH$_3^{12}$C$^{14}$N and H$^{12}$C$^{14}$N from a pure CH$_3^{12}$C$^{14}$N ice irradiated at 15 K. The spectra display the same energy dependence as the TEYs, with a dominant contribution at 400 eV, corresponding to the N~1s~$\rightarrow \pi^*$ resonance of CH$_3$CN. This confirms that the X-ray photodesorption is well correlated to the X-ray photoabsorption of the ice.

\begin{figure*}[!htbp]
\centering
\includegraphics[scale=0.65]{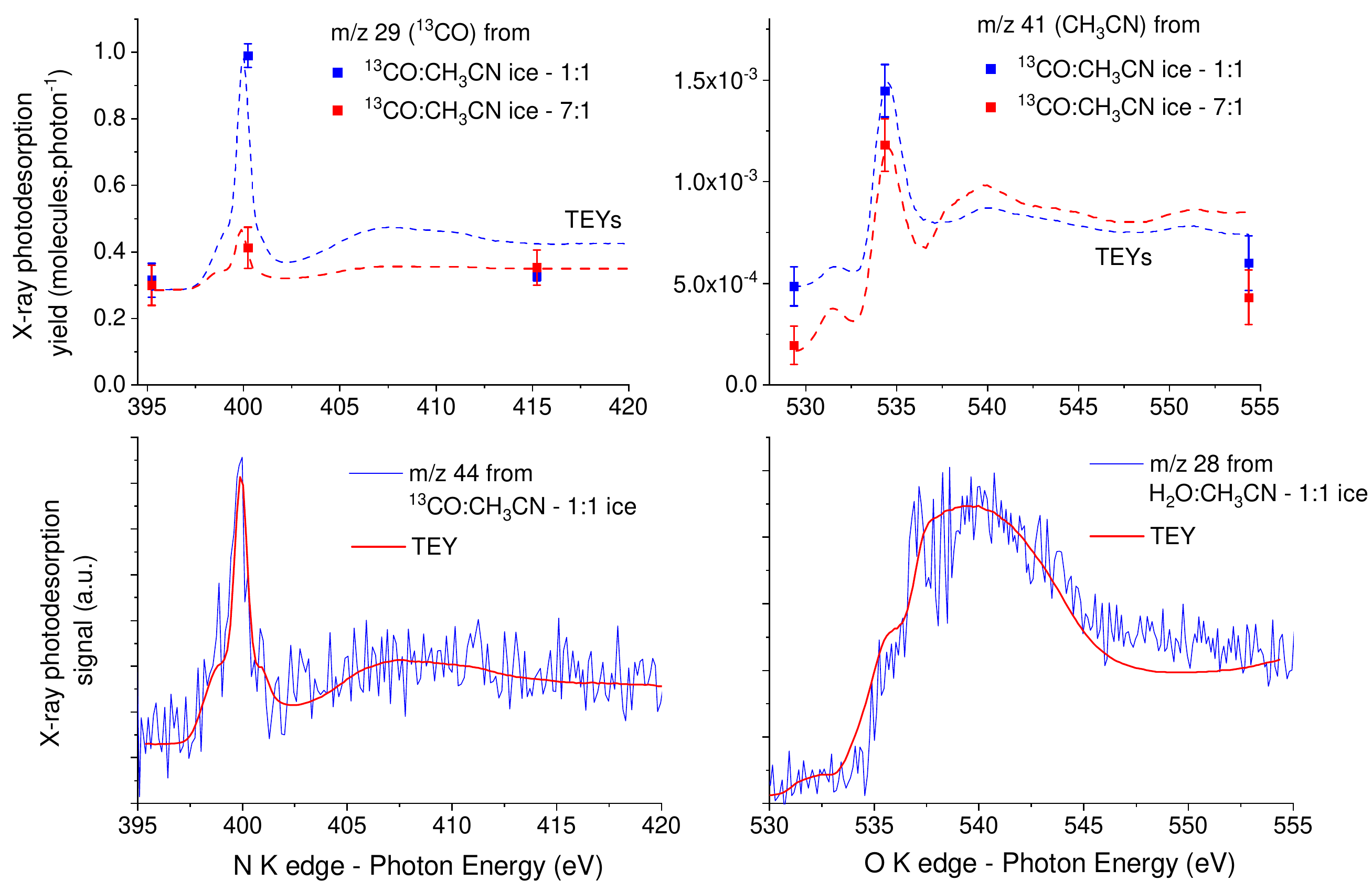} 
\caption{X-ray photodesorption spectra from mixed ices. Top panels: X-ray photodesorption yields of $^{13}$CO near the N K edge and of CH$_3$CN near the O K edge (squares with error bars) from a mixed $^{13}$CO${:}$CH$_3$CN ice at 15 K, with dilution ratios of 1${:}$1 (blue) and 7${:}$1 (red). The yields were measured for a fresh ice at 15 K having received a photon fluence < 1 $\times$ 10$^{16}$ photons.cm$^{-2}$. The TEYs, measured at higher fluences (5 - 10 $\times$ 10$^{16}$ photons.cm$^{-2}$), are displayed as dashed lines. Bottom panels: X-ray photodesorption spectra (blue) of the m/z 44 from a $^{13}$CO${:}$CH$_3$CN (1${:}$1) ice near the N K edge (left panel) and of the m/z 28 from a H$_2$O${:}$CH$_3$CN (1${:}$1) ice near the O K edge (right panel). In red we also display the TEYs. The photodesorption signals and the TEYs were measured simultaneously at 15 K. The Y-scale of the bottom panels is in arbitrary units.}
\label{fig:psd_ch3cn_x_mix}
\end{figure*}

\subsection{X-ray photodesorption from mixed ices}\label{sec:result_mixed}

\begin{figure}[t!]
\centering
\includegraphics[scale=0.47]{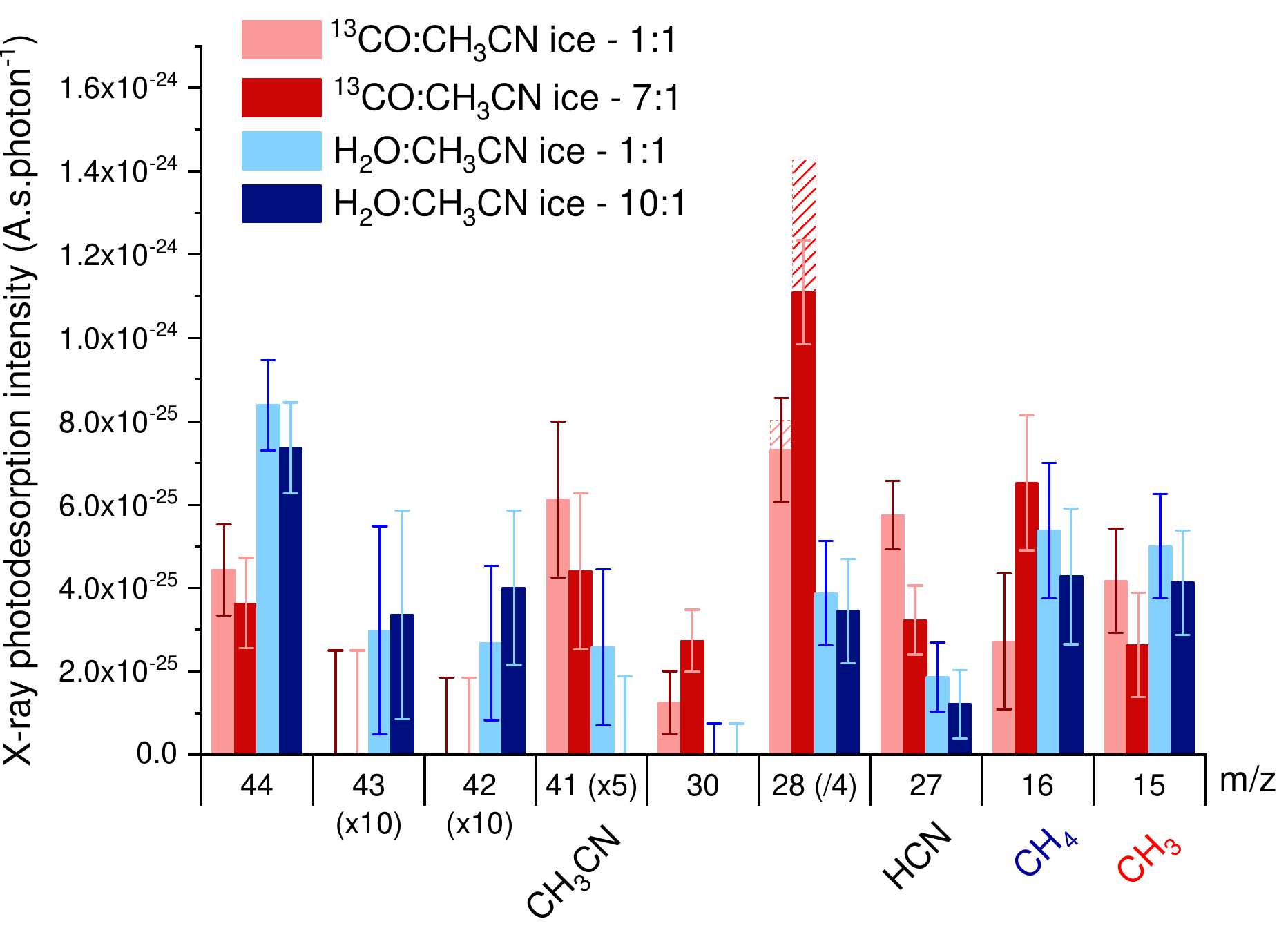} 
\caption{X-ray photodesorption intensities divided by the photon flux (in A.s.photon$^{-1}$) at 560 eV of desorbing masses from mixed $^{13}$CO${:}$CH$_3$CN and H$_2$O${:}$CH$_3$CN ices, irradiated at 15 K. The attribution of the desorbing species to the mass channels is displayed for the m/z 41, 27, 16, and 15. For the other mass channels, this attribution depends on the ice composition. This is discussed in the text. The signals were obtained for a fluence $<$ 2 $\times$ 10$^{16}$ photons.cm$^{-2}$ and are not corrected for any possible fragmentation of desorbing species in the ionization stage of our QMS. The striped rectangles displayed on the m/z 28 for the $^{13}$CO-mixed ices correspond to the contribution of desorbing $^{12}$C$^{16}$O originating from the isotopic impurities in our $^{13}$CO gas sample, as estimated in the text.}
\label{fig:x_ch3cn_attr_mass_mix}
\end{figure}

X-ray irradiation experiments at 15 K in the N and O K edge regions were conducted when CH$_3$CN was mixed in $^{13}$CO or H$_2$O ices with different dilution factors (with a total ice thickness of $\sim$ 100 ML). In these experiments, the acetonitrile isotopolog we used was the natural one ($^{12}$CH$_3$$^{12}$C$^{14}$N). Tuning the photon energy to the N K edge region results in the dominant photoexcitation of CH$_3$CN, whereas tuning the photon energy to the O K edge region results in the dominant photoexcitation of $^{13}$CO or H$_2$O. Examples of X-ray photodesorption spectra from the mixed ices are shown in Figure \ref{fig:psd_ch3cn_x_mix}. In the top panels we display the X-ray photodesorption yields of the m/z 29, which we attribute to $^{13}$CO desorption in the N K edge region (395 - 420 eV), and of the m/z 41, which we attribute to CH$_3$CN desorption in the O K edge region (525 - 555 eV) from $^{13}$CO${:}$CH$_3$CN ices. The variations in photodesorption yields with the photon energy follow that of the TEYs (displayed as dashed lines), that is, the photoabsorption spectrum of the ice. This behavior indicates an indirect photodesorption mechanism in the sense that the photodesorbing molecule is different from the photoexcited one. The top panels of Figure \ref{fig:psd_ch3cn_x_mix} clearly show that photoexciting CH$_3$CN or $^{13}$CO in the N and O K edge region induces the desorption of $^{13}$CO and CH$_3$CN, respectively, from the mixed $^{13}$CO:CH$_3$CN ices.
\\\\
Indirect desorption mechanisms induced by X-ray irradiation of ices have already been highlighted in similar experiments on methanol-containing ices \citep{basalgete_complex_2021, basalgete_complex_2021-b} and on CO/N$_2$ ices \citep{basalgete_2022}. In these studies, it was suggested that the indirect desorption is driven by the scattering of the Auger electrons and the subsequent low-energy secondary electrons toward the ice surface, following the Auger decay of the core hole excited or ionized state of the photoabsorbing molecule. This mechanism, known as X-ray induced electron stimulated desorption (XESD), was also proposed to occur for pure ices of H$_2$O \citep{dupuy_x-ray_2018} and CO \citep{dupuy_x-ray_2021}. We also expect this mechanism to explain the X-ray photodesorption of the neutral species detected in our experiments with acetonitrile-containing ices. Other indirect mechanisms could include the codesorption of molecules at the ice surface, which is not necessarily induced by the secondary electrons, but by the fate of the photoexcited molecule after Auger decay.
\\\\
Additionally, the scattering of the Auger and secondary electrons induces chemistry near the ice surface. The X-ray photodesorption of masses associated with photoproducts were observed during our experiments. Some examples are shown in the  bottom panels of Figure \ref{fig:psd_ch3cn_x_mix} for the X-ray photodesorption signals on the m/z 44 and m/z 28 from a $^{13}$CO:CH$_3$CN ice (1:1) and a H$_2$O:CH$_3$CN ice (1:1), respectively. The fact that the X-ray photodesorption spectra of these masses follow the TEY indicates that these photoproducts originate from the chemistry induced by the low-energy electrons. More globally, when a photodesorption signal was clearly detected during the experiments, the corresponding photodesorption spectrum was found to follow the TEY of the ice.
\\\\
Many mass channels displayed a desorption signal during the experiments performed on mixed ices. In order to discuss the attribution of the neutral species to these signals, we show in Figure \ref{fig:x_ch3cn_attr_mass_mix} the X-ray photodesorption intensities associated with the mass channels we monitored. The displayed intensities are not corrected for any possible fragmentation pattern of the desorbing species. They were derived at 560 eV energy, at which the photoabsorption is dominated by the core ionization of O-bearing species, with a similar cross section for $^{13}$CO and H$_2$O. For similar dilution ratios, potential differences observed in the desorption intensities at 560 eV can therefore be solely attributed to differences in the ice composition. Additionally, the intensities displayed in Figure \ref{fig:x_ch3cn_attr_mass_mix} were obtained for a low photon fluence ($<$ 2 $\times$ 10$^{16}$ photons.cm$^{-2}$) in order to limit the destruction effects of CH$_3$CN before the measurements as much as possible. This is particularly efficient in H$_2$O-dominated ices, as explained in Section \ref{sec:TEY}.
\\\\
The m/z 16 and 15 signals depend on the ice composition. For the $^{13}$CO-mixed ices, the m/z 16 intensity increases with the amount of $^{13}$CO that is initially deposited and it is below our detection limit after correcting it for the fragmentation of desorbing $^{13}$CO into atomic O at the QMS entrance for both mixtures. Consequently, we attribute the intensities we observed on the m/z 15 on the $^{13}$CO-mixed ices to CH$_3$ desorption. For the H$_2$O-mixed ices, the intensities observed on the m/z 16 and 15 are consistent with the desorption of CH$_4$, which should produce a similar signal on these mass channels due to its fragmentation at the QMS entrance. After the cracking of m/z 16 (CH$_4$) into m/z 15 (CH$_3$) was corrected for, the intensities on the m/z 15 from the H$_2$O-mixed ices were not high enough to consider a significant desorption of CH$_3$. We therefore conclude that the m/z 16 and 15 signals from the H$_2$O-mixed ices are solely due to CH$_4$ desorption.
\\\\
The X-ray photodesorption of the m/z 28 and 30 significantly depends on the ice composition. No desorption signal was detected on the m/z 30 from the H$_2$O${:}$CH$_3$CN ices. For the $^{13}$CO-mixed ices, isotopic impurities present in our $^{13}$C$^{16}$O gas sample contribute to the desorption signals observed on the m/z 28 and 30. Mass signals on the m/z 28 and 30 were found in the mass spectrum of our $^{13}$C$^{16}$O gas sample, which was measured after the synchrotron experiments. These mass signals originate from  $^{12}$C$^{16}$O and $^{12}$C$^{18}$O in the gas sample. It was estimated from the mass spectrum that $\sim$ 1\% of $^{12}$C$^{16}$O and $\sim$ 0.5\% of $^{12}$C$^{18}$O relative to $^{13}$C$^{16}$O were present in our sample. In the X-ray photodesorption experiments from the $^{13}$CO-mixed ices, a significant desorption signal was detected on the m/z 29, which is attributed to the X-ray photodesorption of $^{13}$C$^{16}$O. The intensities measured on the m/z 29 and 30 from the $^{13}$CO-mixed ices both increase from the 1:1 to the 7:1 mixtures. Moreover, the ratio of the m/z 30 intensity to that of the m/z 29 was found to be $\sim$ 0.5\% for both mixtures, which is similar to the estimated amount of $^{12}$C$^{18}$O impurities present in our $^{13}$C$^{16}$O gas sample. We therefore conclude that the desorption signal observed on the m/z 30 from the $^{13}$CO-mixed ices solely originates from $^{12}$C$^{18}$O X-ray photodesorption due to the isotopic impurity deposited with the $^{13}$C$^{16}$O matrix.

\ \\
Unlike the m/z 30, the desorption intensities measured on the m/z 28 from the $^{13}$CO-mixed ices cannot be only attributed to the desorption of the $^{12}$C$^{16}$O isotopic impurity deposited in the $^{13}$C$^{16}$O matrix. The part of the m/z 28 desorption intensity attributed to the $^{12}$C$^{16}$O impurity desorption is estimated as 1\% of the $^{13}$C$^{16}$O (m/z 29) desorption intensity, and it is displayed in the striped rectangles in Figure \ref{fig:x_ch3cn_attr_mass_mix}. A significant part of the m/z 28 intensity should therefore originate from the desorption of photoproducts. For both the $^{13}$CO-mixed and H$_2$O-mixed ices, the high desorption intensity as compared to the other mass channels indicates photoproducts with a high desorption efficiency, such as $^{12}$C$^{16}$O or N$_2$. Their X-ray photodesorption yields from their pure respective ice are found to be similar, and they are $\sim$ 0.1 molecules.photon$^{-1}$ (at 420 eV for N$_2$ and 560 eV for CO; \cite{dupuy_x-ray_2021, basalgete_2022}). The accumulation of CO near the ice surface, which should be visible in the TEY data near 534.4 eV, is not observed in the mixed H$_2$O${:}$CH$_3$CN ices (see the top panel of Figure \ref{fig:tey_ch3cn_O}), whereas N$_2$ accumulation is clearly seen in the TEY near 401 eV, even in the low-fluence regime at which the intensities were derived (see the right panels of Figure \ref{fig:tey_ch3cn_N}). This might indicate that the desorption of the m/z 28 from mixed H$_2$O${:}$CH$_3$CN ices is dominated by the X-ray photodesorption of N$_2$, although a contribution of CO desorption cannot be totally ruled out. For the $^{13}$CO${:}$CH$_3$CN ices, the accumulation of N$_2$ is also visible in the TEY near the N K edge, but the formation of $^{12}$CO cannot be discussed regarding the TEY near the O K edge because $^{13}$CO and $^{12}$CO core absorption features should be similar in that energy range. Therefore, we cannot clearly attribute desorbing species to the m/z 28 signals for the mixed ices. It is unclear why this signal increases with the dilution of CH$_3$CN in the $^{13}$CO-mixed ices but is similar in the H$_2$O-mixed ices for both dilution ratios. 
\\\\
The m/z 27 intensities observed for mixed $^{13}$CO${:}$CH$_3$CN and H$_2$O${:}$CH$_3$CN ices decrease when CH$_3$CN is more diluted. This indicates the X-ray photodesorption of HCN. This decrease is somewhat consistent with the fact that this molecule might desorb following the dissociation of CH$_3$CN at the ice surface, for example, via DEA due to low-energy electrons \citep{bass_reactions_2012}. This monomolecular process should not be hindered by the surrounding $^{13}$CO or H$_2$O molecules, but the associated desorption signal should be less detectable when fewer CH$_3$CN molecules are present at the ice surface. A lower desorption signal of HCN for the case of H$_2$O${:}$CH$_3$CN ices compared to the $^{13}$CO${:}$CH$_3$CN ices with similar dilution ratios might be due to (i) a higher desorption barrier to overcome in the case of mixed H$_2$O${:}$CH$_3$CN ices, or (ii) the fact that CH$_3$CN is more rapidly consumed hence present in smaller numbers at the ice surface for the case of  mixed H$_2$O${:}$CH$_3$CN ices, even for the low-fluence regime in which the signals were measured. 
\\\\
The desorption intensities of the m/z 41 are attributed to CH$_3$CN X-ray photodesorption. In mixed $^{13}$CO${:}$CH$_3$CN ices, it decreases when CH$_3$CN is more diluted. The signals in the H$_2$O${:}$CH$_3$CN ices are lower than in the $^{13}$CO${:}$CH$_3$CN ices for similar dilution ratios, indicating that mixing CH$_3$CN with water tends to hinder its X-ray photodesorption. As for the case of HCN, this could be due to either a difference in adsorption energies between CH$_3$CN and the water or the $^{13}$CO matrix, or it could be due to a rapid consumption of CH$_3$CN by X-ray induced chemistry when it is mixed with water. Desorption signals on higher m/z than 41, namely 42, 43, and 44, were observed from the mixed ices. The desorbing species cannot be clearaly attributed to these mass channels because these mass channels can correspond to $^{12}$CNO, H$^{12}$CNO, and $^{12}$CO$_2$ for the m/z 42, 43, and 44, respectively, or they can originate from the fragmentation of molecules of higher mass at the QMS entrance. The relative intensity of these signals between the $^{13}$CO-dominated and H$_2$O-dominated ice indicates a different X-ray induced chemistry. H$^{12}$CNO desorption from the H$_2$O-mixed ices and further fragmentation at the QMS entrance might explain the signals observed on the m/z 42 and 43. The relative intensities of these signals are not consistent with the mass spectrum measured in \cite{Hand_1971}, but the large uncertainties make it difficult to firmly exclude HCNO. Its formation in our H$_2$O-mixed ices is supported by the detection of its conjugated base, OCN$^-$, in 0.8 MeV proton and far-UV irradiated H$_2$O:CH$_3$CN ices in the study of \cite{HUDSON2004466}. In the presence of water, taking inspiration from the H$^+$ assisted hydrolysis of nitrile, the induced chemistry might result in the conversion of the nitrile function into an amide group R-CO-NH$_2$. Part of the m/z 44, whose global intensity is doubled in the case of the mixed H$_2$O:CH$_3$CN ices compared to the mixed $^{13}$CO:CH$_3$CN ices, can be attributed to the carbamoyl-ionized radical [H$_2$N-CO]$^+$, for example, resulting from the cracking of acetamide CH$_3$-CONH$_2$. Another possibility to explain the m/z 44 signal from the H$_2$O-mixed ices would be the desorption and further fragmentation of formamide HCONH$_2$. Formamide desorption should produce a signal on the m/z 45 that is higher than the m/z 43 and 44 according to the NIST database \citep{NIST_chemistry_webbook}, however. As we did not detect a desorption signal on the m/z 45 from the H$_2$O-mixed ices, we might exclude formamide desorption even if the large uncertainties prevent us from concluding this definitively. Both acetamide and formamide have been proposed as possible photoproducts in the VUV photolysis of H$_2$O:CH$_3$CN ices \citep{bulak_photolysis_2021}.
\\\\
After conversion of the desorption intensities into desorption yields, we display in Table \ref{tab:yield_ch3cn_x_astro-fragments} the X-ray photodesorption yields at 560 eV that are associated with the identified species from the mixed ices, with dilution ratios that are the most representative of interstellar ices, that is the higher ratios.
{\renewcommand{\arraystretch}{1.2}
\begin{table}[t!]
\begin{center}
\caption{\label{tab:yield_ch3cn_x_astro-fragments} X-ray photodesorption yields in molecules desorbed per incident photon (in 10$^{-4}$ molecules.photon$^{-1}$) of CH$_3$CN, HCN, CH$_4$ , and CH$_3$ at 560 eV from mixed $^{13}$CO${:}$CH$_3$CN and H$_2$O${:}$CH$_3$CN ices irradiated at 15 K. }
\begin{tabular}{p{1.8cm}p{2.5cm}p{2.8cm}}
 \hline \hline
 Species & CO${:}$CH$_3$CN - 7${:}$1 & H$_2$O${:}$CH$_3$CN - 10${:}$1   \\\hline
CH$_3$CN & 4.3 $\pm$ 1.2  & < 1 \\
HCN & 8.3 $\pm$ 2.1 &  3.1 $\pm$ 2.1 \\
CH$_4$ & < 2 & 6.8 $\pm$ 2.7   \\
CH$_3$ & 3.9 $\pm$ 2.3 & < 2 \\
 \hline \hline
 \end{tabular}
\end{center}
\tablefoot{The fluence received by the ice before the measurements is $<$ 2 $\times$ 10$^{16}$ photons.cm$^{-2}$}
\end{table}
}

\section{Astrophysical yields and discussion}\label{sec:astro}
The most important finding of this study is that X-ray photodesorption of intact CH$_3$CN from interstellar ices is a possible process and might partly explain the observation of CH$_3$CN in protoplanetary disks \citep{oberg_comet-like_2015, bergner_survey_2018, loomis_distribution_2018}. According to our experimental results, the efficiency of this process should depend on the ice composition and hence on the disk region that is considered. Namely, in regions in which CH$_3$CN is mixed in CO-dominated ices at the ice surface, the X-ray photodesorption yield of CH$_3$CN is expected to be higher than the one corresponding to the regions in which CH$_3$CN is mixed in H$_2$O-dominated ices at the ice surface. X-ray photodesorption of photofragments, for instance, HCN, CH$_4$ , and CH$_3$, should also enrich the gas phase of disks with these molecules. The effect of the ice temperature on the X-ray photodesorption yields remains to be studied because it varies with the regions of the disk that are considered, with H$_2$O ices being warmer than CO ices. Although it was shown that X-ray irradiation alone can promote diffusion of species in ices \citep{Jiménez-Escobar_2022}, the increase in ice temperature should also favor the diffusion of photoproducts and might influence the photodesorption. Due to the indirect desorption processes observed in our experiments, most probably mediated by the Auger scattering and the subsequent cascade of low-energy secondary electrons, the X-ray photoabsorption of any subsurface molecule in interstellar ices should induce the desorption of surface molecules. The ice depth involved in this indirect process is expected to be a few tens of ML in the soft X-ray range, based on similar X-ray experiments \citep{basalgete_2022}.   
{\renewcommand{\arraystretch}{1.4}
\begin{table*}[t!]
\begin{center}
\caption{\label{tab:yield_astro}X-ray astrophysical photodesorption yields in molecules desorbed per incident photon (molecules.photon$^{-1}$) of CH$_3$CN, HCN, CH$_4$ , and CH$_3$ from CO-dominated and H$_2$O-dominated ices at 15 K. }
\begin{tabular}{p{3cm}p{3cm}p{3cm}p{3cm}p{3cm}}
 \hline \hline
 & CH$_3$CN & HCN & CH$_4$ & CH$_3$  \\\hline
 Mixed with CO \\\hline
 Source Spectrum & 2.0$^{\pm 1.0} \times$ 10$^{-4}$ & 3.9$^{\pm 2.0} \times$ 10$^{-4}$ & < 1.0$^{\pm 0.5} \times$ 10$^{-4}$ & 1.8$^{\pm 0.9} \times$ 10$^{-4}$ \\
 $N_H$ = 10$^{21}$ cm$^{-2}$ & 1.8$^{\pm 0.9} \times$ 10$^{-4}$ & 3.6$^{\pm 2.8} \times$ 10$^{-4}$ & < 9.4$^{\pm 4.7} \times$ 10$^{-5}$ & 1.7$^{\pm 0.8} \times$ 10$^{-4}$ \\
 $N_H$ = 10$^{22}$ cm$^{-2}$ & 9.1$^{\pm 4.5} \times$ 10$^{-5}$ & 1.8$^{\pm 0.9} \times$ 10$^{-4}$ & < 4.3$^{\pm 2.1} \times$ 10$^{-5}$ & 8.2$^{\pm 4.1} \times$ 10$^{-5}$ \\
 $N_H$ = 10$^{23}$ cm$^{-2}$ & 1.2$^{\pm 0.6} \times$ 10$^{-5}$ & 2.3$^{\pm 1.2} \times$ 10$^{-5}$ & < 5.6$^{\pm 2.8} \times$ 10$^{-6}$ & 1.1$^{\pm 0.6} \times$ 10$^{-5}$ \\
 $N_H$ = 10$^{24}$ cm$^{-2}$ & 1.6$^{\pm 0.8} \times$ 10$^{-6}$ & 3.1$^{\pm 1.5} \times$ 10$^{-6}$ & < 7.5$^{\pm 3.7} \times$ 10$^{-7}$ & 1.4$^{\pm 0.7} \times$ 10$^{-6}$ \\\hline
 Mixed with H$_2$O \\\hline
 Source Spectrum & < 5.6$^{\pm 2.8} \times$ 10$^{-5}$ & 1.5 $^{\pm 0.7} \times$ 10$^{-4}$ & 3.2$^{\pm 1.6} \times$ 10$^{-4}$ & < 1.1$^{\pm 0.6} \times$ 10$^{-4}$ \\
 $N_H$ = 10$^{21}$ cm$^{-2}$ & < 5.1$^{\pm 2.5} \times$ 10$^{-5}$ & 1.4$^{\pm 0.7} \times$ 10$^{-4}$ & 3.0$^{\pm 1.5} \times$ 10$^{-4}$ & < 1.0$^{\pm 0.5} \times$ 10$^{-4}$ \\
 $N_H$ = 10$^{22}$ cm$^{-2}$ & < 2.4$^{\pm 1.2} \times$ 10$^{-5}$ & 7.2$^{\pm 3.6} \times$ 10$^{-5}$ & 1.6$^{\pm 0.8} \times$ 10$^{-4}$ & < 4.7$^{\pm 2.3} \times$ 10$^{-5}$  \\
 $N_H$ = 10$^{23}$ cm$^{-2}$ & < 3.2$^{\pm 1.6} \times$ 10$^{-6}$ & 9.8$^{\pm 4.9} \times$ 10$^{-6}$ & 2.2$^{\pm 1.1} \times$ 10$^{-5}$ & < 6.4$^{\pm 3.2} \times$ 10$^{-6}$ \\
 $N_H$ = 10$^{24}$ cm$^{-2}$ & < 4.3$^{\pm 2.6} \times$ 10$^{-7}$ & 1.3$^{\pm 0.6} \times$ 10$^{-6}$ & 2.9$^{\pm 1.4} \times$ 10$^{-6}$ & < 8.6$^{\pm 4.3} \times$ 10$^{-7}$  \\
 \hline \hline
 \end{tabular}
\end{center}
\tablefoot{These yields were derived according to the method described in Section \ref{sec:exper}.}
\end{table*}
}
\ \\\\
In order to provide quantitative data that could be easily implemented in astrochemical models, we derive in Table \ref{tab:yield_astro} astrophysical yields according to the method described in Section \ref{sec:exper}, that is, by extrapolation of the experimental yields in the 0.4 - 10 keV and by averaging them on the estimated local X-ray spectrum, which depends on the column density of gas and dust $N_H$ traversed by the stellar X-rays. The X-ray emission spectrum is that of a typical T-Tauri star, taken from \cite{nomura_molecular_2007}, and the attenuation cross section of gas and dust is taken from \cite{bethell_photoelectric_2011}. We consider in Table \ref{tab:yield_astro} the case where CH$_3$CN is diluted in a CO-dominated or H$_2$O-dominated ice, which corresponds to the higher dilution ratios studied in our experiments (7:1 for the CO-dominated ice and 10:1 for the H$_2$O-dominated ice). For a given ice composition, the astrophysical yields vary by two orders of magnitude, depending on the local X-ray spectrum and hence on the disk region that is considered. 
\\\\
For regions in which hard X-rays (> 1 keV) dominate the spectrum (see Figure \ref{fig:fig_local_phi} of Appendix \ref{App:astro}), the photodesorption yields are lowest. This is due to our extrapolation that results in yields that are several orders of magnitude lower for hard X-rays than for soft X-rays. This extrapolation represents a non-negligible uncertainty on the yields displayed in Table \ref{tab:yield_astro}, and additional experiments should be conducted in the hard X-ray range to estimate its accuracy. It might indeed be expected that for these energies, the X-ray photoabsorption results in the scattering of both the primary ionized 1s electron and the Auger electron toward the ice surface, inducing desorption. As the kinetic energy of the 1s electron increases with the X-ray energy, a deviation of the photodesorption yields from the photoabsorption cross section might be observed for hard X-rays.    
\\\\
For CO-dominated ices and for a similar attenuation of the X-ray irradiation spectrum, the estimated astrophysical X-ray photodesorption yields of CH$_3$CN are found to be in the same order of magnitude than that of another important COM in astrochemistry, methanol CH$_3$OH \citep{basalgete_complex_2021-b}. Moreover, the X-ray photodesorption behavior of the intact COM is found to be similar between CH$_3$CN-containing ices and CH$_3$OH-containing ices \citep{basalgete_complex_2021-b}. The X-ray photodesorption yield of the intact COM, either CH$_3$CN or CH$_3$OH, is estimated to be lower in the case of H$_2$O-dominated ices than in the case of CO-dominated ices. For both COMs, this is assumed to be due to a difference in the X-ray induced chemistry between H$_2$O-dominated and CO-dominated ices: in the water matrix, chemical reactions between the intact COM and most probably OH radical tend to increase the destruction kinetic of the COM, which competes with its intact desorption. Interestingly, experiments in the VUV range on CH$_3$CN-containing ices and CH$_3$OH-containing ices display a different behavior of the intact COM desorption. For CO-dominated ices, the VUV photodesorption of CH$_3$CN ($\sim$ 10$^{-5}$ molecules.photon$^{-1}$ at 10.5 eV, from \cite{basalgete_photodesorption_2021}) is found to be at least an order of magnitude higher than that of CH$_3$OH (only an upper limit of $\sim$ 10$^{-6}$ molecules.photon$^{-1}$ has been derived in \cite{bertin_uv_2016}). Additionally, the VUV photodesorption of CH$_3$CN is found to be independent of the studied ice composition (pure CH$_3$CN ices, CO:CH$_3$CN ices, or H$_2$O:CH$_3$CN ices; \cite{basalgete_photodesorption_2021}), in contrast to what is observed in the X-ray range in our study. This shows that X-ray and VUV photodesorption of COMs should not be treated similarly in astrochemical models as very different physical-chemical mechanisms are expected to be at play for these processes. Finally, it is not straightforward to conclude on the dominant role of either VUV photons or X-rays for the photodesorption of COMs in protoplanetary disks by solely considering our experimental data. Our X-ray yields are found to significantly depend on the disk region considered, and they are found to be either superior or inferior to the the VUV yields by an order of magnitude. For CH$_3$CN, the maximum astrophysical yield derived, which is $\sim$ 10$^{-4}$ molecules.photon$^{-1}$, is still an order of magnitude lower than what is used in the study of \cite{loomis_distribution_2018}.
In order to easily extrapolate our experimental results to environments other than protoplanetary disks, we also provide in Table \ref{tab:yield_absorbed_photon} the X-ray photodesorption yields in units of absorbed photons by using the yields at 560 eV from Table \ref{tab:yield_ch3cn_x_astro-fragments}. 
{\renewcommand{\arraystretch}{1.2}
\begin{table}[b!]
\begin{center}
\caption{\label{tab:yield_absorbed_photon} X-ray photodesorption yields in 10$^{-2}$ molecules desorbed per absorbed photon of CH$_3$CN, HCN, CH$_4$ , and CH$_3$ from mixed $^{13}$CO${:}$CH$_3$CN and H$_2$O${:}$CH$_3$CN ices irradiated at 15 K. }
\begin{tabular}{p{1.8cm}p{2.5cm}p{2.8cm}}
 \hline \hline
 Species & CO${:}$CH$_3$CN & H$_2$O${:}$CH$_3$CN  \\\hline
CH$_3$CN & 2.4 $\pm$ 1.2  & < 0.5 \\
HCN & 4.5 $\pm$ 2.2 &  1.6 $\pm$ 0.8 \\
CH$_4$ & < 1 & 3.6 $\pm$ 1.8\\
CH$_3$ & 2.1 $\pm$ 1.0 & < 1 \\
 \hline \hline
 \end{tabular}
\end{center}
\tablefoot{These yields were derived from those displayed in Table \ref{tab:yield_ch3cn_x_astro-fragments} at 560 eV and assuming that up to 30 ML of the ice contribute to the desorption.}
\end{table}
}

\section{Conclusion}
X-ray photodesorption of neutral species from CH$_3$CN-containing ices was studied in the soft X-ray range in the N and O K edge regions (395- 420 eV and 530 - 555 eV, respectively). X-ray photodesorption yields of CH$_3$CN, HCN, CH$_4$ , and CH$_3$ were derived for pure CH$_3$CN ice, $^{13}$CO:CH$_3$CN ices, and H$_2$O:CH$_3$CN ices. The yields were found to depend on the photon energy and on the ice composition. Indirect desorption processes, induced by photoexcitation of either CH$_3$CN, $^{13}$CO, or H$_2$O, and most probably mediated by the Auger scattering and the subsequent cascade of low-energy electrons, were observed. The X-ray photodesorption yield at 560 eV of the intact CH$_3$CN was estimated to be higher by at least half an order of magnitude when CH$_3$CN is mixed in CO-dominated ices compared to the case where it is mixed in H$_2$O-dominated ices. 
\\\\
X-ray photodesorption of intact CH$_3$CN from interstellar ices might partly explain the abundances of gas-phase CH$_3$CN observed in protoplanetary disks. The desorption efficiency depends on the local X-ray irradiation spectrum and on the ice composition and hence on the disk region that is considered. In order to facilitate the implementation of X-ray photodesorption in disk modeling, we derived astrophysical yields, averaged in the 0.4 - 10 keV range, as a function of the local conditions expected in disks. For the desorption of the intact CH$_3$CN, these astrophysical yields vary from $\sim$ 10$^{-4}$ molecules.photon$^{-1}$ to $\sim$ 10$^{-6}$ molecules.photon$^{-1}$ from CO-dominated ices. Only upper limits, from $\sim$ 5 $\times$ 10$^{-5}$ molecules.photon$^{-1}$ to $\sim$ 5 $\times$ 10$^{-7}$ molecules.photon$^{-1}$, could be derived for the X-ray photodesorption of CH$_3$CN from H$_2$O-dominated ices.

\begin{acknowledgements}
This work was carried out with financial support from the Region Ile-de-France DIM-ACAV + program; the Sorbonne Université “Emergence” program; the ANR PIXyES project, Grant No. ANR-20-CE30-0018 of the French “Agence Nationale de la Recherche”; and the Program National “Physique et Chimie du Milieu Interstellaire” (PCMI) of CNRS/INSU with INC/INP cofunded by CEA and CNES. We would like to acknowledge SOLEIL for the provision of synchrotron radiation facilities under Project No. 20210142 and N. Jaouen, H. Popescu, and R. Gaudemer for their help on the SEXTANTS beam line.
\end{acknowledgements}

%
\bibliographystyle{aa} 
\bibliography{MyBibli} 
%

\begin{appendix} 

\section{Raw desorption mass signals}
 \label{App:app_A}
 
\begin{figure}[b!]
   \centering
   \includegraphics[scale=0.5]{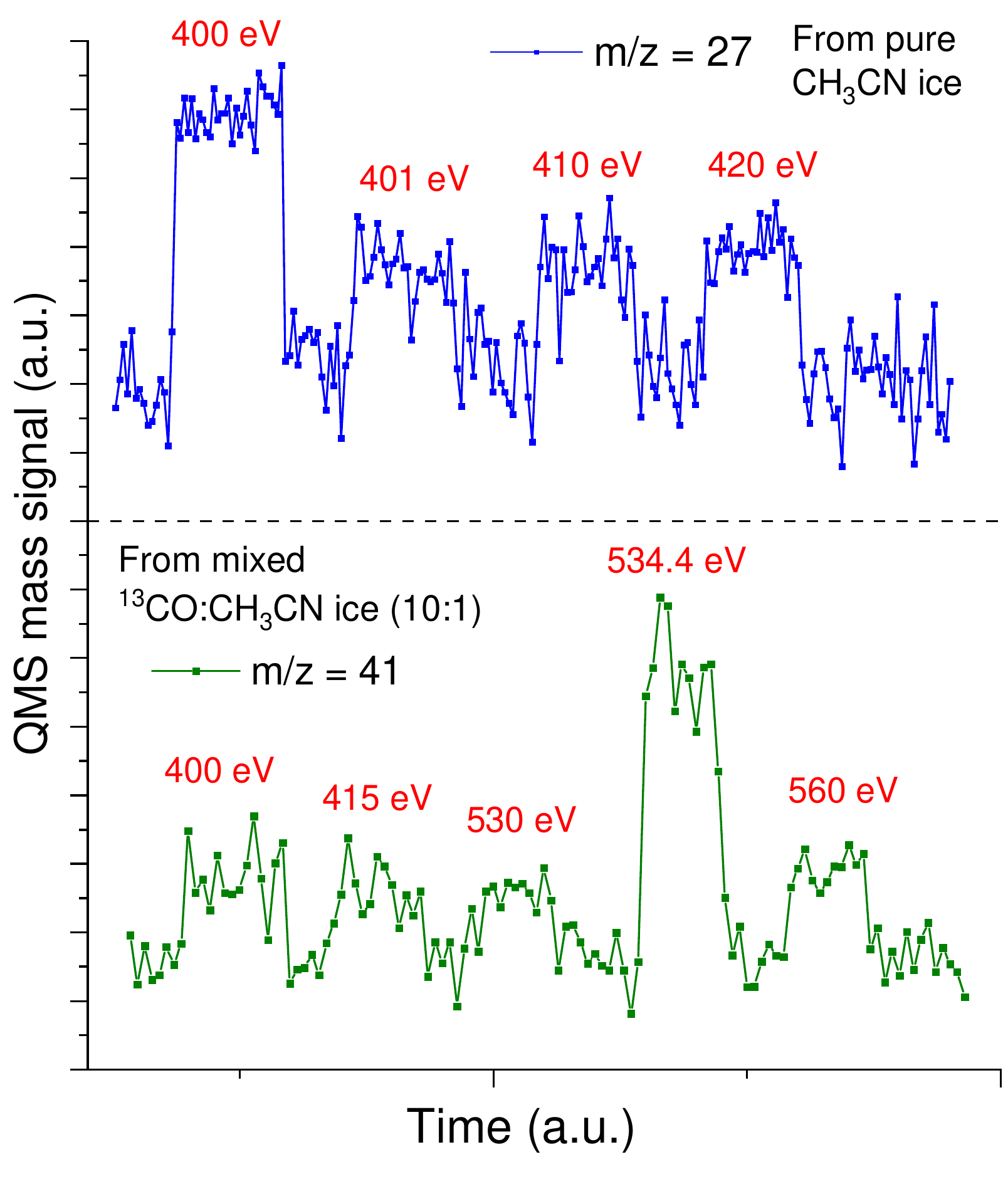}
   \caption{X-ray photodesorption QMS signals of m/z 27 from pure CH$_3$CN ice (top part) and of m/z 41 from mixed $^{13}$CO:CH$_3$CN (10:1) ice (bottom part), both irradiated at 15 K and at a fixed photon energy for a few tens of seconds. The fixed energies at which the ice is irradiated are indicated in red for each irradiation step. The QMS signals are scaled for more clarity, meaning that the relative intensities between the m/z 27 and 41 signals do not represent the actual ones.}
              \label{fig:fig_raw_data_1}%
\end{figure}

\begin{figure}[t!]
   \centering
   \includegraphics[scale=0.75]{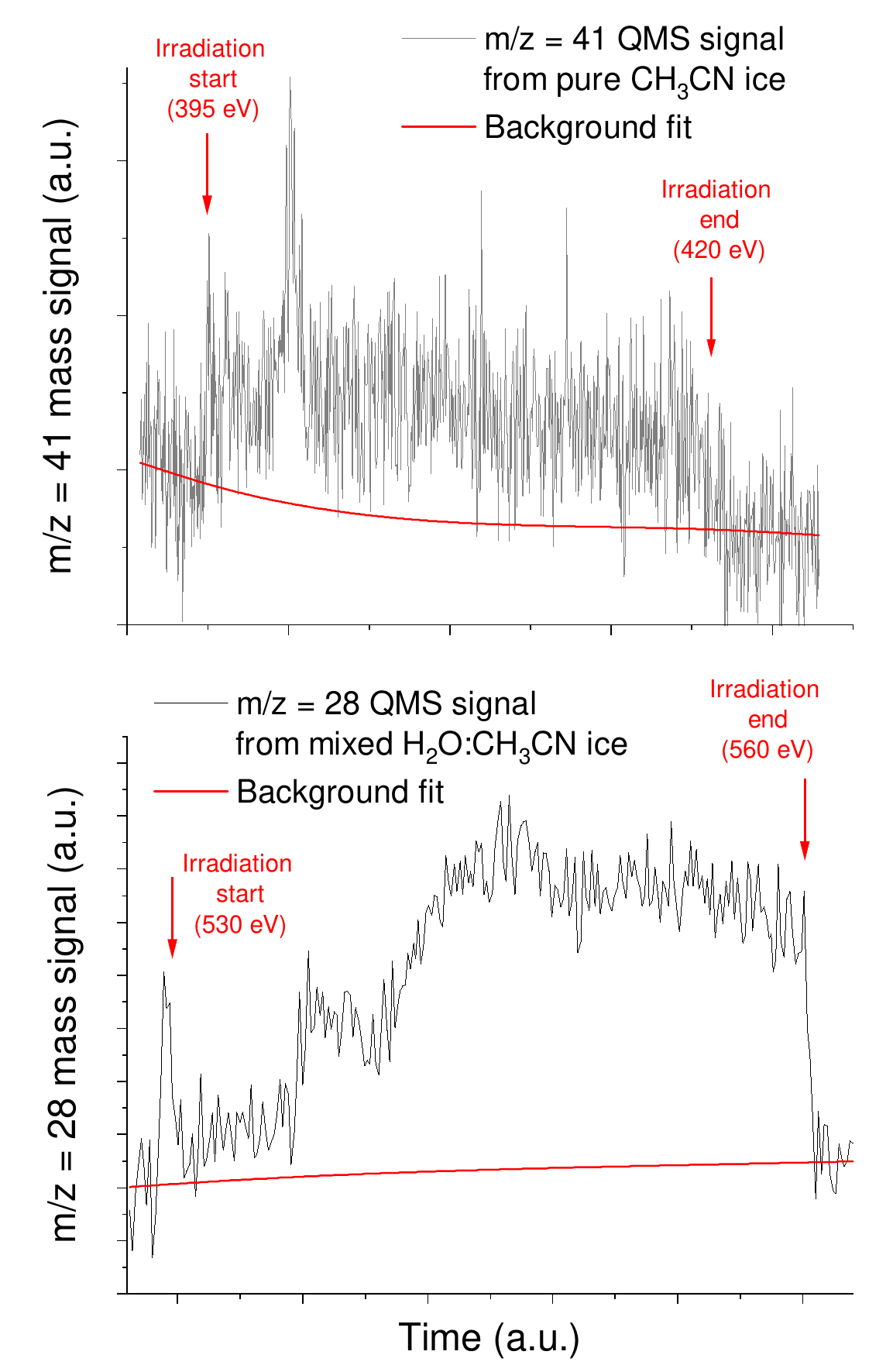}
   \caption{Examples of raw photodesorption spectra from the studied ices. Top panel: X-ray photodesorption QMS signal of m/z 41 from pure CH$_3$CN ice irradiated at 15 K by scanning the incident photon energy between 395 eV and 420 eV. Bottom panel: X-ray photodesorption QMS signal of m/z 28 from mixed H$_2$O:CH$_3$CN (1:1) ice irradiated at 15 K by scanning the incident photon energy between 530 eV and 560 eV. The start and the end of the irradiation are shown by the red arrows.}
              \label{fig:fig_raw_data_2}%
\end{figure}

 In Figure \ref{fig:fig_raw_data_1} and \ref{fig:fig_raw_data_2} we display some examples of QMS signals obtained during the experiments we conducted. As explained in the text, the desorption intensities $I_X(E)$ are deduced from these data either by computing the signal height (with respect to the background level) corresponding to the irradiation step at fixed energy (Figure \ref{fig:fig_raw_data_1}) or by subtracting the background fit from the continuous mass signal corresponding to an irradiation scan (Figure \ref{fig:fig_raw_data_2}). 
  
\ \\
The X-ray photodesorption yields displayed in Tables \ref{tab:yield_ch3cn_x_pure} and \ref{tab:yield_ch3cn_x_astro-fragments} are derived from the fixed energy experiments. The uncertainties shown in these tables and the error bars also displayed in Figures \ref{fig:fig_psd_pure_ch3cn} and \ref{fig:psd_ch3cn_x_mix} are associated with the signal-to-noise ratio of the QMS mass signals (see Figure \ref{fig:fig_raw_data_1}). First, the uncertainties associated with $I_X(E)$, $\delta I_X(E)$, are estimated as the sum of the half-height of the signal noise in the background and at the top of each peak for the corresponding irradiation step. Then, for each energy and each species, the uncertainties associated with the X-ray photodesorption yields $\delta \Gamma_X(E)$, which are displayed in the mentioned tables and figures, are deduced from the following equation: 
\begin{equation}
    \delta \Gamma_X(E) = k_X \frac{\delta I_X(E)}{\phi(E)}
    \label{eq:app_a}
.\end{equation}
Equation \ref{eq:app_a} results in experimental uncertainties that depend on the photon energy. Finally, these uncertainties only reflect the quality of the recorded signals and do not take the uncertainty associated with the calibration of $k_X$ into account, which is systemic. These uncertainties do not affect the shape of the X-ray photodesorption spectra, but only the uncertainties on the absolute values of the photodesorption yields. It mainly depends on the uncertainties associated with the apparatus function of the QMS and on the electron-impact ionization cross sections. We estimate these latter uncertainties to be in the order of magnitude of 50\% relative to the derived photodesorption yields.

\section{Astrophysical X-ray photodesorption yields}\label{App:astro}

\begin{figure}[t!]
\centering
\includegraphics[scale=1.2]{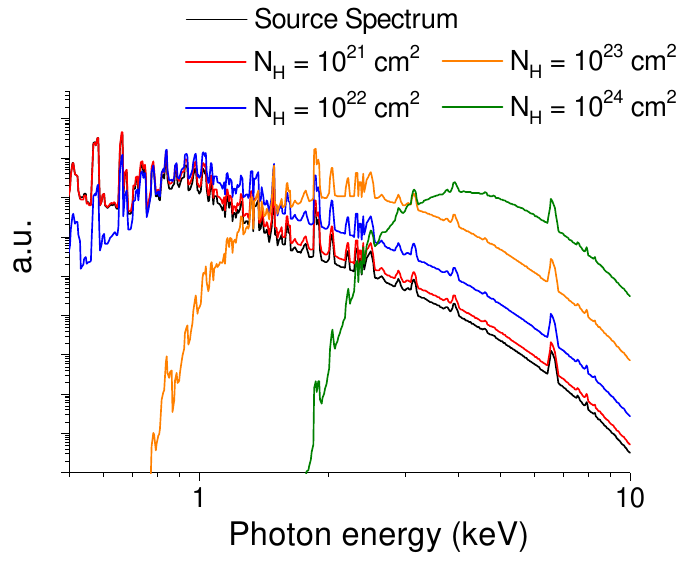} 
\caption{Normalized source and attenuated X-ray spectrum of a typical T-Tauri star. The source spectrum is taken from \cite{nomura_molecular_2007}.}
\label{fig:fig_local_phi}
\end{figure}
\begin{figure}[b!]
\centering
\includegraphics[scale=0.65]{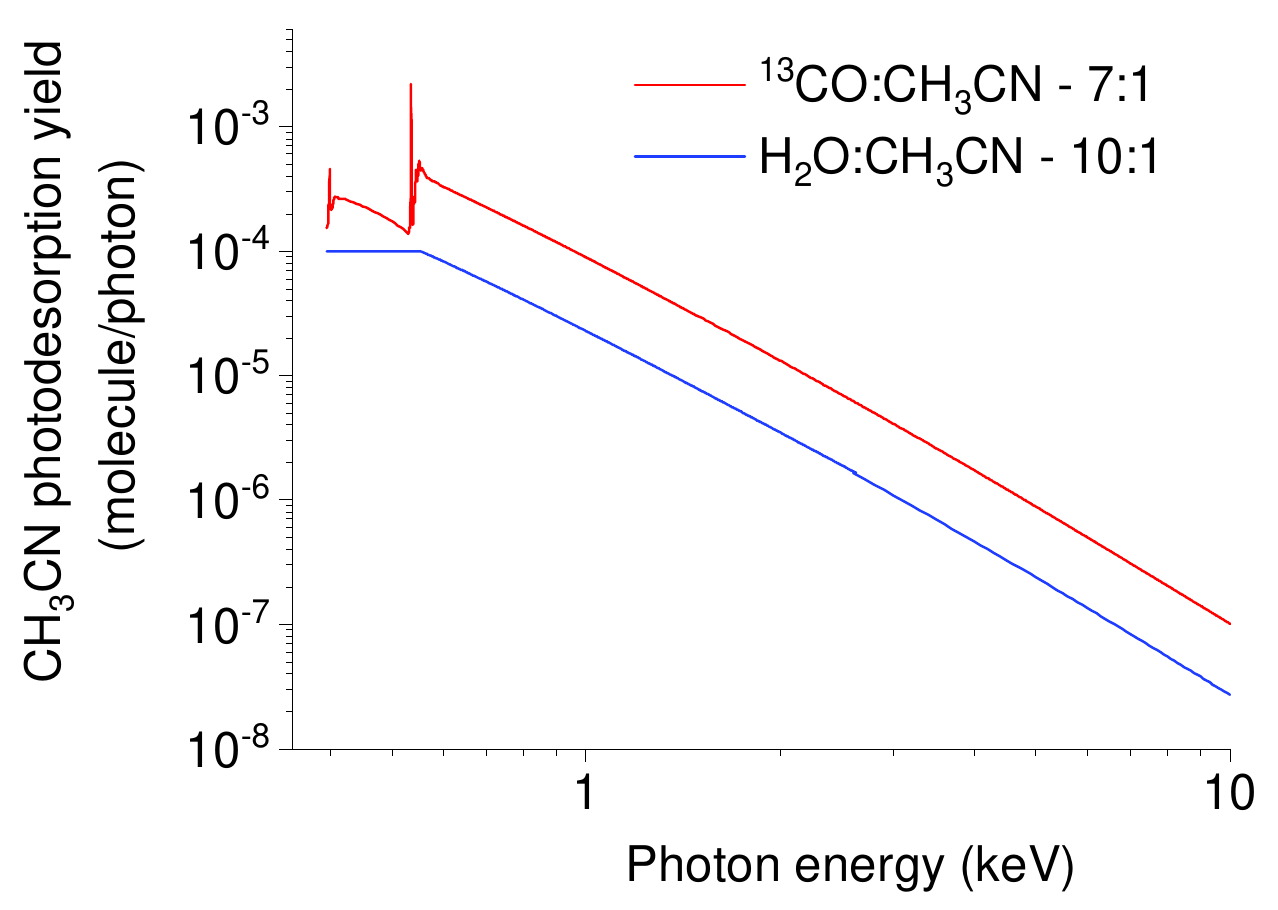} 
\caption{X-ray photodesorption yields of CH$_3$CN from mixed $^{13}$CO:CH$_3$CN (10:1) and H$_2$O:CH$_3$CN (10:1) ices at 15 K, extrapolated from 0.4 to 10 keV according to the methodology described in Section \ref{sec:exper}.}
\label{fig:fig_extrapol}
\end{figure}

In order to derive X-ray photodesorption yields averaged over the 0.4 - 10 keV range, referred to as astrophysical yields, we used the method described in Section \ref{sec:exper} that is summarized in equation \ref{eq:yield_astro}. In this equation, $\phi_{local}$ is estimated by taking the X-ray emission source spectrum of a typical T-Tauri star $\phi_{source}$ (taken from \cite{nomura_molecular_2007}) and by attenuating it using the photoelectric cross section $\sigma_{att}$ of gas and dust from \cite{bethell_photoelectric_2011}. This results in the following formula:
\begin{equation}
    \phi_{local}(E) = \phi_{source}(E) \ e^{- \sigma_{att}(E) \ N_H}
,\end{equation}
where $N_H$ is the column density of gas and dust traversed by the X-rays. The corresponding local X-ray spectra $\phi_{local}$, expected to irradiate the interstellar ices, depend on the disk region and are displayed in Figure \ref{fig:fig_local_phi}. 
\\\\
Equation \ref{eq:yield_astro} necessitates extrapolating the experimental X-ray photodesorption yields $\Gamma_X$, derived in Section \ref{sec:results}, to higher X-ray energies. As explained in Section \ref{sec:exper}, this extrapolation consists of assuming that the yields follow the O 1s ionization cross section of CO (or similarly H$_2$O) taken from \cite{Berkowitz:1087021}. As the X-ray photodesorption of CH$_3$CN from a mixed H$_2$O:CH$_3$CN (10:1) ice is below our detection limit, we considered for the extrapolation that the CH$_3$CN yield was in that case equal to our upper limit, which is 10$^{-4}$ molecules.photon$^{-1}$, in the 395 - 560 eV range. The extrapolated CH$_3$CN X-ray photodesorption yields from mixed $^{13}$CO:CH$_3$CN (10:1) and H$_2$O:CH$_3$CN (10:1) are displayed in Figure \ref{fig:fig_extrapol}

\section{Total electron yields near the O K edge with photon fluence}\label{App_B} 
\begin{figure}[h!]
\centering
\includegraphics[scale=0.58]{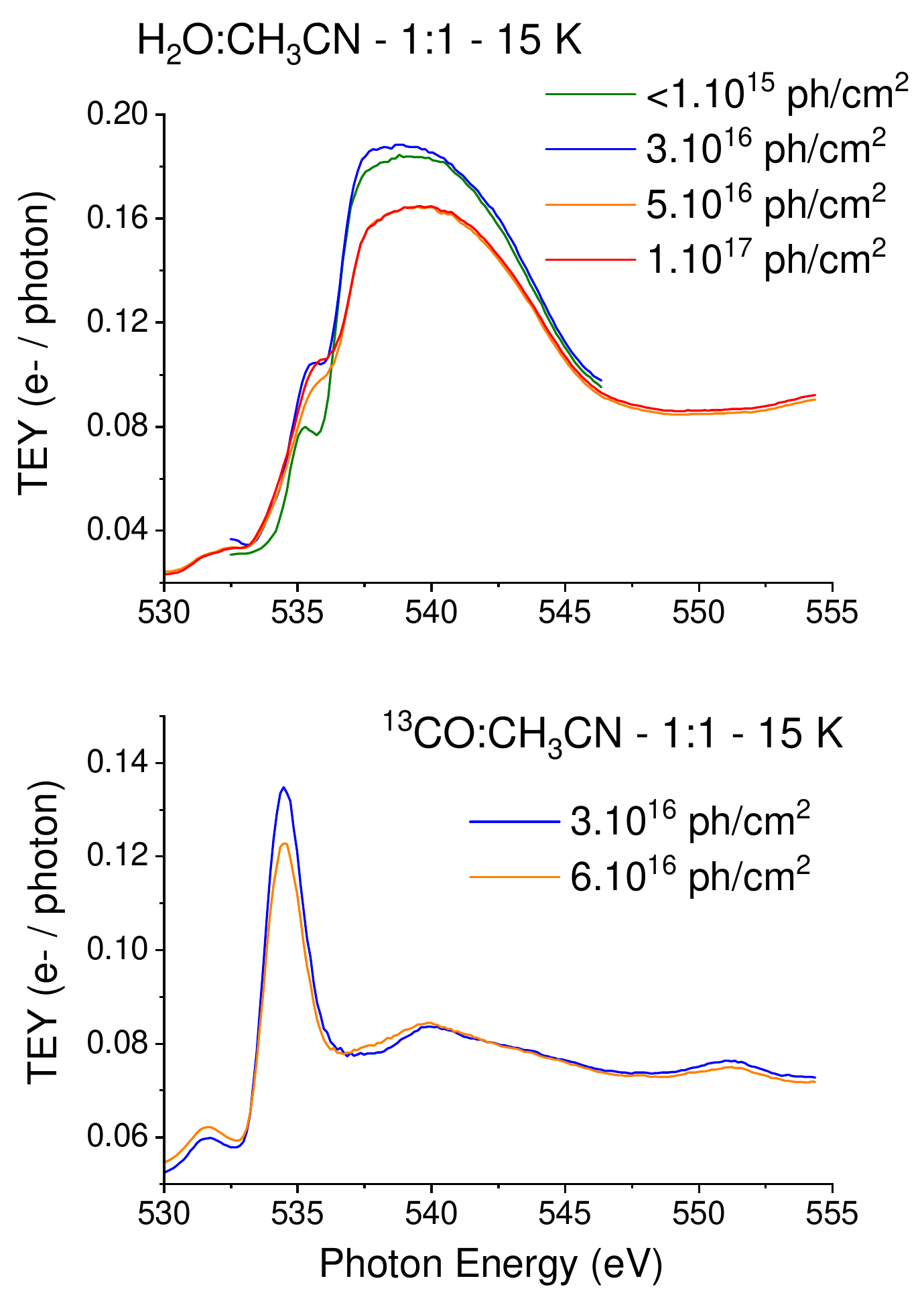} 
\caption{TEYs near the O K edge of a mixed H$_2$O${:}$CH$_3$CN ice irradiated at 15 K with a dilution ratio of 1${:}$1 (top panel) and of a mixed $^{13}$CO${:}$CH$_3$CN ice irradiated at 15 K with a dilution ratio of 1${:}$1 (bottom panel). The photon fluence received by the ice before the TEY measurement is also displayed.}
\label{fig:tey_ch3cn_O}
\end{figure}

In Figure \ref{fig:tey_ch3cn_O} we display the TEYs measured from mixed H$_2$O${:}$CH$_3$CN and $^{13}$CO${:}$CH$_3$CN ices irradiated at 15 K near the O K edge as a function of the photon fluence. The features observed are similar to that corresponding to pure H$_2$O and pure CO ice that were studied in \cite{dupuy_desorption_2020} and \cite{dupuy_x-ray_2021}. As explained in the text, no significant new features are detected in these TEYs data, showing that potential photoproducts that formed during X-ray irradiation do not strongly participate in the ice absorption in this energy range. The evolution of the feature observed near 535.4 eV with the photon fluence for the mixed H$_2$O${:}$CH$_3$CN ice is most likely due to a slight change in the structure of the water ice. As discussed in \cite{dupuy_desorption_2020}, this feature can be associated with single-donor weakly coordinated H$_2$O molecules in the water ice, which might increase after the formation of defects due to X-ray irradiation at 15 K. 

\end{appendix}

\end{document}